\documentclass[reprint,twocolumn,aps,superscriptaddress]{revtex4-1}
\usepackage[utf8]{inputenc}
\usepackage[T1]{fontenc}
\usepackage{amsmath}
\usepackage{amsfonts}
\usepackage{physics}
\usepackage[caption=false]{subfig}
\usepackage{bm}
\usepackage{amssymb}
\usepackage[mathscr]{eucal}
\usepackage{graphicx}
\usepackage[dvipsnames]{xcolor}
\usepackage[hidelinks,colorlinks=true,allcolors=Blue]{hyperref}
 
\begin{document}
\title{Molecular and solid-state topological polaritons induced by population imbalance}
\author{Sindhana Pannir-Sivajothi}
\affiliation{Department of Chemistry and Biochemistry, University of California
San Diego, La Jolla, California 92093, USA}
\author{Nathaniel P. Stern}
\affiliation{Department of Physics and Astronomy, Northwestern University, Evanston, Illinois 60208, USA}
\author{Joel Yuen-Zhou}
\email{joelyuen@ucsd.edu}
\affiliation{Department of Chemistry and Biochemistry, University of California
San Diego, La Jolla, California 92093, USA}
\begin{abstract}
Strong coupling between electronic excitations in materials and photon modes results in the formation of polaritons, which display larger nonlinearities than their photonic counterparts due to their material component. We theoretically investigate how to optically control the topological properties of molecular and solid-state exciton-polariton systems by exploiting one such nonlinearity: saturation of electronic transitions. We demonstrate modification of the Berry curvature of three different materials when placed within a Fabry-Perot cavity and pumped with circularly polarized light, illustrating the broad applicability of our scheme. Importantly, while optical pumping leads to non-zero Chern invariants, unidirectional edge states do not emerge in our system as the bulk-boundary correspondence is not applicable. This work demonstrates a versatile approach to control topological properties of novel optoelectronic materials.
\end{abstract}
\maketitle

\section*{Introduction}
\begin{figure}[b]
	\centering
	\includegraphics[width=0.7\columnwidth]{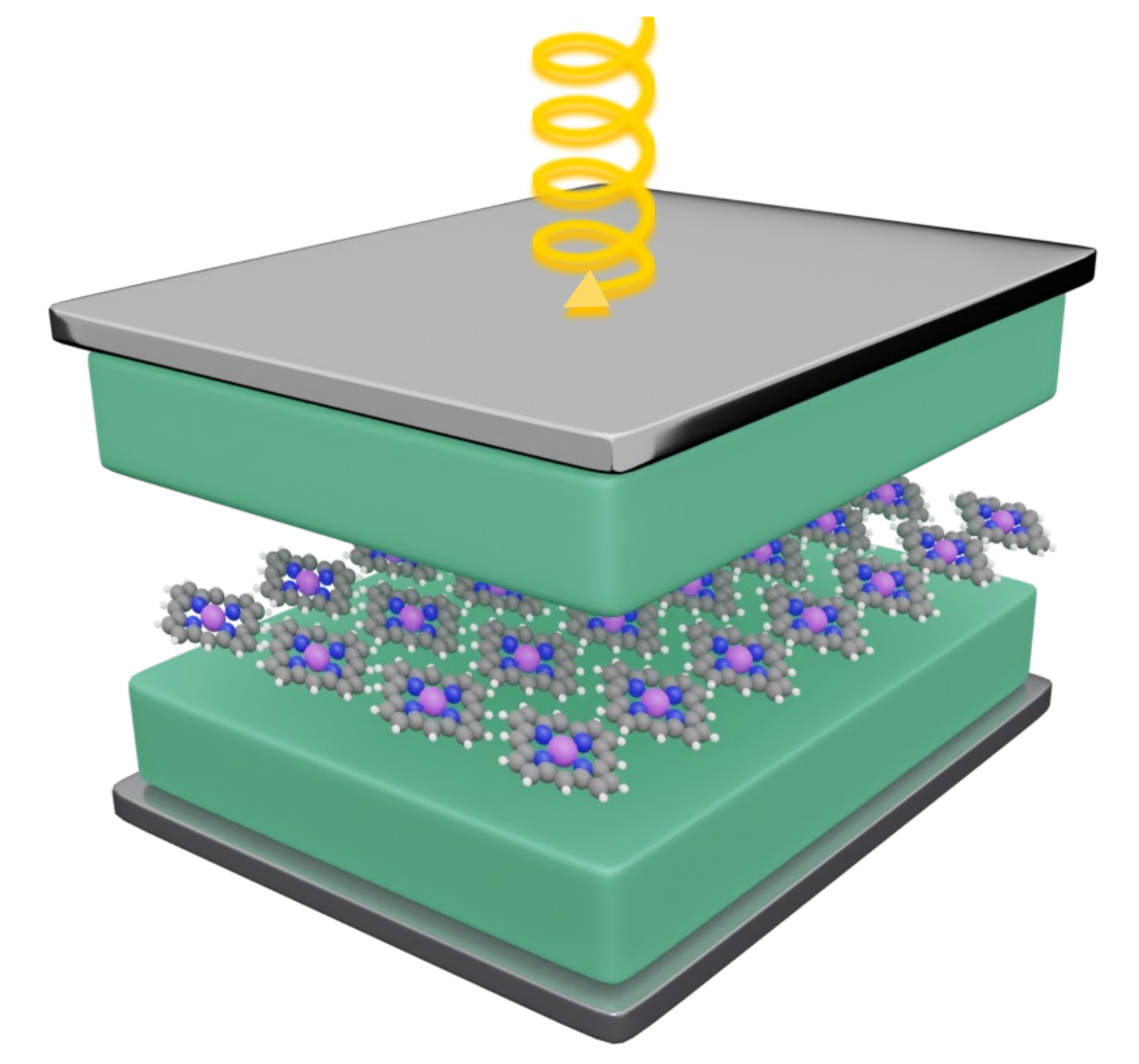}
	\caption{\label{fig:intro}Illustration of the system under study. Porphyrin (molecules at the center) and perylene (green blocks) placed within a Fabry-Perot cavity and pumped with circularly polarized light.}
\end{figure}

Exciton-polaritons are hybrid excitations that exist in systems where photonic modes couple strongly with optical transitions in materials and their coupling strength exceeds losses \cite{weisbuch1992observation}. Electronic strong coupling (ESC), where the optical transitions correspond to semiconductor excitons or molecular electronic transitions, has been observed in a wide variety of inorganic and organic materials. While some polariton systems, such as GaAs and CdTe quantum wells in microcavities \cite{weisbuch1992observation,andre1998spectroscopy}, often require cryogenic temperatures for operation, due to their small exciton binding energies, organic materials \cite{lidzey1998strong} along with others such as GaN \cite{butte2006room}, ZnO \cite{shimada2008cavity}, perovskites \cite{brehier2006strong,su2021perovskite}, and transition metal dichalcogenides (TMD) \cite{liu2015strong,hu2020recent} can achieve ESC at room temperature when placed in Fabry-Perot cavities. In particular, organic exciton-polaritons have received attention for their ability to modify chemical reactivity \cite{hutchison2012modifying}, demonstrate polariton condensation at room temperature \cite{daskalakis2014nonlinear,dietrich2016exciton}, improve photoconductivity \cite{krainova2020polaron}, and display topological properties \cite{liao2021experimental,dusel2021room}.

Exciton-polariton systems are versatile platforms for topological applications as their hybrid nature provides the unique opportunity to take advantage of the nonlinearities and magnetic response of the material component while still enjoying benefits of the coherence properties of the photonic part \cite{solnyshkov2021microcavity,bardyn2015topological,yuen2016plexciton}. In the presence of photonic lattices, they also offer the possibility of unidirectional transport of energy through edge states that are robust to disorder \cite{klembt2018exciton}. A few approaches are frequently used to achieve topological exciton-polariton bands. 
In one of the approaches, the non-trivial topology resides in the winding light-matter coupling rather than individual photon or exciton components \cite{karzig2015topological,klembt2018exciton}. However, it is limited in application due to the requirement of large magnetic fields to break time-reversal symmetry (TRS) and low temperatures to achieve Zeeman splitting in the exciton component which exceeds the exciton linewidth. In another approach, TRS is preserved and a quantum spin hall insulator analogue is created in a polariton system \cite{liu2020generation}. This approach does not require a large magnetic field, however, there, a topological polariton system is created by coupling a  topologically non-trivial photonic lattice with a topologically trivial exciton system and the interesting topology is almost entirely encoded in the photonic component of the polariton \cite{liu2020generation,li2021experimental}. Both the approaches mentioned above were experimentally realized in polariton lattices. More recently, polaritons in  Fabry-Perot cavities have emerged as a viable platform for topological polaritonics. Several experiments have demonstrated measurement and control of the Berry curvature of exciton-polariton and photon bands in these systems \cite{gianfrate2020measurement,rechcinska2019engineering,ren2021nontrivial,lempicka2022electrically}. Our work will focus on these Fabry-Perot cavity systems.

In this work, we theoretically propose a scheme for generating topological polaritons that combines advantages of both the approaches mentioned above. Specifically, we exploit the primary nonlinearity of organic exciton-polaritons, saturation \cite{daskalakis2014nonlinear}, to achieve this. Here, the light-matter coupling contains the non-trivial topology instead of the individual photon or exciton components and optical pumping with circularly polarized light breaks TRS instead of a large magnetic field.

Breaking TRS in a system using the helicity of light is an idea that has been demonstrated in several other contexts; it has been used to achieve all-optical non-reciprocity \cite{guddala2021all,lenferink2014coherent} and theoretical results suggest that it can also induce optical-activity in achiral molecules \cite{schwennicke2020optical}. Additionally, a similar idea that relies on breaking TRS using circularly polarized light has been previously proposed for polariton lattices by Bleu \textit{et al.} \cite{bleu2017photonic}.

We focus on the topological properties of polaritons formed by the coupling of Frenkel excitons hosted in organic semiconductors with photon modes in a Fabry-Perot cavity. Here, optical pumping with circularly polarized light saturates certain electronic transitions and breaks TRS in the system; this results in non-zero Chern numbers of polariton bands. Our scheme relies on the contraction of Rabi splitting due to saturation, and we find modified Berry curvature and Chern number of the bands under circularly polarized pumping. The Berry curvature of the more photonic sections of the bands computed in our work can be experimentally measured using pump-probe spectroscopy. Furthermore, the applicability of our scheme is not limited to organic polariton systems. It only requires certain key ingredients: transitions that can be selectively excited with circularly polarized light, saturation effects, and Rabi splitting contraction. To highlight this, we compute the Berry curvature of two other systems under strong coupling and optical pumping: (a) Ce:YAG and (b) monolayer MoS$_2$. Our work provides a viable strategy to induce non-reciprocal behavior in standard microcavity polaritons, leading to the optical tuning of isolators and circulators \cite{guddala2021all}, as well as fabrication of elliptically-polarized lasers and condensates \cite{long2022helical}. 

\section*{Results}
\subsection*{Model}
In our theoretical study, we consider a Fabry-Perot cavity containing a thin film of porphyrin molecules at the center and a bulk perylene crystal filling the rest of the volume (Fig. \ref{fig:intro}). The porphyrin and perylene molecules are not treated on an equal footing in our model; while the molecular transitions of porphyrin are considered explicitly in the Hamiltonian, those of the perylene crystal are not, and they can be accounted for through effective cavity modes \cite{ren2021nontrivial}. This is a valid approximation because we focus on photon modes with frequencies close to those of electronic transitions in porphyrin ($\sim3.81$eV) \cite{rubio1999theoretical,edwards1970porphyrins} and far off-resonant from the transitions of perylene ($\sim2.98$eV) \cite{rangel2018low}. Here, the birefringent perylene crystal plays the role of providing anisotropy and emergent optical activity to the cavity modes \cite{ren2021nontrivial}.

\begin{figure}[t]
	\centering
	\includegraphics[width=\columnwidth]{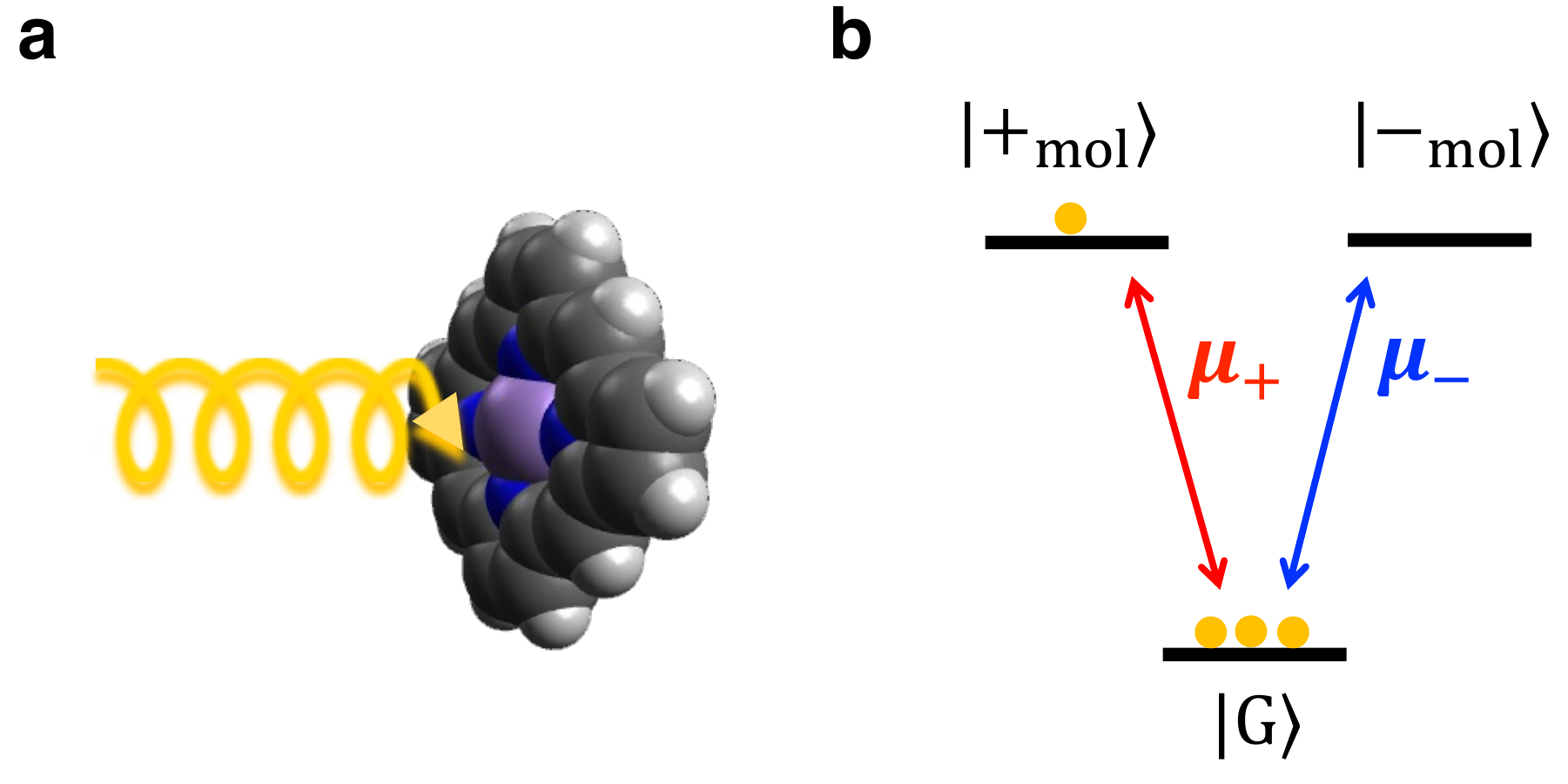}
	\caption{\label{fig:porphyrin3LS}Three-level model of a metalloporphyrin molecule. (a) Illustration of circularly polarized light exciting a metalloporphyrin molecule. (b) Three-level model of porphyrin with a ground state $\ket{\mathrm{G}}$ and two degenerate excited states $\ket{+_{\mathrm{mol}}}, \ket{-_{\mathrm{mol}}}$. The transition dipole moment for a transition from $\ket{\mathrm{G}}$ to $\ket{\pm_{\mathrm{mol}}}$ is $\bm{\mu}_{\pm}=\mu_0(\mathbf{\hat{x}}\pm i\mathbf{\hat{y}})/\sqrt{2}$. The number of yellow circles at each state represents the fraction of molecules in that state. Here, the ratio of the fraction of molecules in the ground, $f_{\mathrm{G}}$, and $\ket{\pm_{\mathrm{mol}}}$ excited states, $f_{\pm}$, is $f_{\mathrm{G}}:f_+:f_-=3:1:0$. Such population ratios can be achieved through pumping with circularly polarized light.}
\end{figure}

We model each porphyrin molecule as a three-level electronic system with a ground state $\ket{\mathrm{G}}$ and two excited states $\ket{+_{\mathrm{mol}}}$ and $\ket{-_{\mathrm{mol}}}$ (see Fig. \ref{fig:porphyrin3LS}b) \cite{barth2006unidirectional,yuen2014topologically}. In the absence of a magnetic field, the two excited states are degenerate and the energy difference between the ground and excited states is $\hbar\omega_{\mathrm{e}}=3.81$eV \cite{sun2022polariton}. The transition dipole moments for transitions from $\ket{\mathrm{G}}$ to $\ket{+_{\mathrm{mol}}}$ and $\ket{-_{\mathrm{mol}}}$ are $\bm{\mu}_{+}=\mu_0(\mathbf{\hat{x}}+i\mathbf{\hat{y}})/\sqrt{2}$ and $\bm{\mu}_{-}=\mu_0(\mathbf{\hat{x}}-i\mathbf{\hat{y}})/\sqrt{2}$, respectively, with $\mu_0=2.84  \mathrm{D}$ \cite{sun2022polariton}. Here, $\mathbf{\hat{x}}$ and $\mathbf{\hat{y}}$ are unit vectors along the x and y directions. Using circular polarized light, the $\ket{+_{\mathrm{mol}}}$ or $\ket{-_{\mathrm{mol}}}$ states can be selectively excited.

In our model, we consider a thin film of metalloporphyrins or metallophtalocyanines arranged in a square lattice with nearest neighbor spacing $a$. The choice of lattice is irrelevant because later we will take the continuum limit $a\to 0$ as we are only interested in length scales much larger than the intermolecular spacing. Additionally, we use periodic boundary conditions along the $x$ and $y$ directions and consider a box of size $L_x\cross L_y$. Each molecule is labeled with the index $\mathbf{m}=(m_{x},m_{y})$ that specifies its location in a $N_x\cross N_y$ array of molecules where $L_{x/y}=aN_{x/y}$; here, the molecule's position is given by $\mathbf{r}_{\mathbf{m}}=m_{x} a\mathbf{\hat{x}}+m_{y} a\mathbf{\hat{y}}$. States of the $\mathbf{m}^{\mathrm{th}}$ molecule are then written as $\ket{\mathbf{m},\mathrm{G}}$, $\ket{\mathbf{m},+_{\mathrm{mol}}}$ and $\ket{\mathbf{m},-_{\mathrm{mol}}}$. The creation operator $\hat{\sigma}^{\dagger}_{\mathbf{m},\pm}=\ket{\mathbf{m},\pm_{\mathrm{mol}}}\bra{\mathbf{m},\mathrm{G}}\otimes_{\mathbf{n}\neq\mathbf{m}}\mathbb{I}_{\mathbf{n}}$ excites the $\mathbf{m}^{\mathrm{th}}$ molecule from $\ket{\mathbf{m},\mathrm{G}}$ to $\ket{\mathbf{m},\pm_{\mathrm{mol}}}$. Here, $\mathbb{I}_{\mathbf{n}}=\ket{\mathbf{n},\mathrm{G}}\bra{\mathbf{n},\mathrm{G}}+\ket{\mathbf{n},+_{\mathrm{mol}}}\bra{\mathbf{n},+_{\mathrm{mol}}}+\ket{\mathbf{n},-_{\mathrm{mol}}}\bra{\mathbf{n},-_{\mathrm{mol}}}$ is the identity operator for $\mathbf{n}^{\mathrm{th}}$ molecule. These molecular operators satisfy commutation relations (a generalization of the commutation relations of paulion operators \cite{mukamel1999principles,agranovich2009excitations}), 
\begin{equation}\label{eq:commutator}
	\comm{\hat{\sigma}_{\mathbf{n},\pm}}{\hat{\sigma}^{\dagger}_{\mathbf{m},\pm}}
	=\delta_{\mathbf{m},\mathbf{n}}(1-\hat{\sigma}_{\mathbf{n},\mp}^{\dagger}\hat{\sigma}_{\mathbf{n},\mp}-2\hat{\sigma}_{\mathbf{n},\pm}^{\dagger}\hat{\sigma}_{\mathbf{n},\pm}).
\end{equation}

We model the effective photon modes of a Fabry-Perot cavity filled with perylene as in Ren \textit{et al}. \cite{ren2021nontrivial} For the photon modes of a Fabry-Perot cavity, the component of wave vector orthogonal to the mirrors $k_z=2n_z\pi/L_z$ is quantized, where $L_z$ is the effective distance between the mirrors of the cavity and $n_z$ is the mode index \cite{kavokin2017microcavities}. For a given $n_z$, the modes are labeled by the in-plane wave vector $\mathbf{k}=k_{x}\mathbf{\hat{x}}+k_{y}\mathbf{\hat{y}}$ and polarization $\alpha$; the creation operators associated with these modes are $\hat{a}_{\mathbf{k},\alpha}^{\dagger}$ and they satisfy bosonic commutation relations $\comm{\hat{a}_{\mathbf{k},\alpha}}{\hat{a}_{\mathbf{k}',\alpha'}^{\dagger}}=\delta_{\alpha,\alpha'}\delta_{\mathbf{k},\mathbf{k}'}$. As a result of in-plane translational invariance of the cavity and periodic boundary conditions along the $x$ and $y$ directions, $k_x=2l_x\pi/L_x$ and $k_y=2l_y\pi/L_y$ take a discrete but infinite set of values $l_x,l_y\in\mathbb{Z}$. Throughout this work, we specify the cavity mode polarization in the circularly polarized basis $\alpha=\pm$.

The Hamiltonian of the full system is
\begin{equation}
	\hat{H}=\hat{H}_{\mathrm{mol}}+\hat{H}_{\mathrm{cav}}+\hat{H}_{\mathrm{cav-mol}},
\end{equation}
where
\begin{equation}
	\begin{aligned}
		\hat{H}_{\mathrm{mol}}=&\sum_{\mathbf{m}}\Big(\hbar\omega_{\mathrm{e}} \hat{\sigma}_{\mathbf{m},+}^{\dagger}\hat{\sigma}_{\mathbf{m},+}+\hbar\omega_{\mathrm{e}}\hat{\sigma}_{\mathbf{m},-}^{\dagger}\hat{\sigma}_{\mathbf{m},-}\Big)\\
		\hat{H}_{\mathrm{cav}}=&\sum_{\mathbf{k}}\Bigg[\Big(E_0+\frac{\hbar^2|\mathbf{k}|^2}{2m^*}+\zeta|\mathbf{k}|\cos\phi\Big)\hat{a}_{\mathbf{k},+}^{\dagger}\hat{a}_{\mathbf{k},+}\\
		&+\Big(E_0+\frac{\hbar^2|\mathbf{k}|^2}{2m^*}-\zeta|\mathbf{k}|\cos\phi\Big)\hat{a}_{\mathbf{k},-}^{\dagger}\hat{a}_{\mathbf{k},-}\\
		&+\Big(-\beta_0+\beta|\mathbf{k}|^2e^{-i2\phi}\Big)\hat{a}_{\mathbf{k},+}^{\dagger}\hat{a}_{\mathbf{k},-}\\
		&+\Big(-\beta_0+\beta|\mathbf{k}|^2e^{i2\phi}\Big)\hat{a}_{\mathbf{k},-}^{\dagger}\hat{a}_{\mathbf{k},+}\Bigg],\\
		\hat{H}_{\mathrm{cav-mol}}=&\sum_{\mathbf{m}}\sum_{\mathbf{k},\alpha}-\bm{\hat{\mu}}_{\mathbf{m}}\cdot\mathbf{\hat{E}}_{\mathbf{k},\alpha}(\mathbf{r}_{\mathbf{m}},0)\\
		\approx&\sum_{\mathbf{m}}\sum_{\mathbf{k}}\frac{e^{i\mathbf{k}\cdot \mathbf{r_m}}}{\sqrt{N_xN_y}}\Bigg[(\bm{\mu}_{+}\cdot\mathbf{J}_{\mathbf{k},+})\hat{\sigma}_{\mathbf{m},+}^{\dagger}\hat{a}_{\mathbf{k},+}\\
		&+(\bm{\mu}_{-}\cdot\mathbf{J}_{\mathbf{k},+})\hat{\sigma}_{\mathbf{m},-}^{\dagger}\hat{a}_{\mathbf{k},+}+(\bm{\mu}_{+}\cdot\mathbf{J}_{\mathbf{k},-})\hat{\sigma}_{\mathbf{m},+}^{\dagger}\hat{a}_{\mathbf{k},-}\\
		&+(\bm{\mu}_{-}\cdot\mathbf{J}_{\mathbf{k},-})\hat{\sigma}_{\mathbf{m},-}^{\dagger}\hat{a}_{\mathbf{k},-}\Bigg]+\mathrm{H.c.}
	\end{aligned}
\end{equation}
Above, $\hat{H}_{\mathrm{mol}}$ describes the porphyrin molecules, $\hat{H}_{\mathrm{cav}}$ the effective cavity modes (including contributions from the perylene crystal), and $\hat{H}_{\mathrm{cav-mol}}$ the coupling between the porphyrin molecules and effective cavity modes. Here, $\phi$ is the angle between the in-plane wave vector and the $x$-axis, \textit{i.e.}, $\cos\phi=k_x/|\mathbf{k}|$. Within $\hat{H}_{\mathrm{cav}}$, $\beta$ specifies the TE-TM splitting, $\beta_0$ quantifies the linear birefringence of the perylene crystal which splits the H-V modes, and $\zeta$ describes the emergent optical activity \cite{ren2021nontrivial}. Additionally, $E_0$ is the frequency of the cavity modes at $|\mathbf{k}|=0$ in the absence of the perylene crystal ($\beta_0=0$ and $\zeta=0$), and $m^*$ is the effective mass of the photons in the absence of perylene ($\beta_0=0$ and $\zeta=0$) and TE-TM splitting ($\beta=0$). We have made the electric dipole approximation and the rotating-wave approximation in $\hat{H}_{\mathrm{cav-mol}}$. Here, $\bm{\hat{\mu}}_{\mathbf{m}}$ is the electric dipole operator associated with the $\mathbf{m}^{\mathrm{th}}$ molecule and $\mathbf{\hat{E}}_{\mathbf{k},\alpha}(\mathbf{r},z)$ is the electric field operator of the mode with polarization $\alpha$ and in-plane wave vector $\mathbf{k}$. In addition, $\bm{\mu}_{\alpha'}\cdot\mathbf{J}_{\mathbf{k},\alpha}$ is the collective coupling strength of the cavity mode labeled by $\mathbf{k},\alpha$ and the $\ket{\mathrm{G}}$ to $\ket{\alpha'_{\mathrm{mol}}}$ transition of the molecules (see Section S1 in Supporting Information). In the Hamiltonian, we only include cavity modes with $\mathbf{k}$ that lies within the first Brillouin zone determined by the porphyrin lattice $-\pi/a<k_x,k_y<\pi/a$. We ignore cavity modes with larger wavevectors (Umklapp terms) as they are off-resonant and would have a negligible effect on the bands of our interest.

The photon modes of an empty cavity experience TE-TM splitting due to polarization dependent reflection from the mirrors \cite{panzarini1999exciton}. While the TE-TM splitting lifts the degeneracy between photon modes at $|\mathbf{k}|\neq0$, photon modes of both polarizations remain degenerate at $|\mathbf{k}|=0$ due to rotational symmetry of the cavity mirrors about the z-axis. However, for Berry curvature and Chern invariant to be well-defined, we need the photon/polariton bands to be separated in energy at all $\mathbf{k}$; to achieve this, we include the perylene crystal. The anisotropy and emergent optical activity of the perylene crystal lifts the degeneracy between the photon modes at all $\mathbf{k}$ \cite{ren2021nontrivial}.

To compute the Berry curvature and Chern number, we focus on the first excitation manifold which is spanned by states $\ket{\mathbf{m},\pm_{\mathrm{mol}}}=\hat{\sigma}_{\mathbf{m},\pm}^{\dagger}\ket{\mathrm{vac}}$ and $\ket{\mathbf{k},\pm_{\mathrm{cav}}}=\hat{a}_{\mathbf{k},\pm}^{\dagger}\ket{\mathrm{vac}}$. Here, $\ket{\mathrm{vac}}$ is the absolute ground state of the system where the photon modes are empty and all molecules are in their ground states.
Rewriting the Hamiltonian with operators $\hat{\sigma}_{\mathbf{k},\alpha}$, where $\hat{\sigma}_{\mathbf{m},\alpha}=\frac{1}{\sqrt{N_xN_y}}\sum_{\mathbf{k}\in\mathrm{BZ}}e^{i\mathbf{k}\cdot\mathbf{r}_{\mathbf{m}}}\hat{\sigma}_{\mathbf{k},\alpha}$ and restricting ourselves to the first excitation manifold, we find $\hat{H}(\mathbf{k})=\bra{\mathbf{k}}\hat{H}\ket{\mathbf{k}}$ to be
\begin{equation}
	\hat{H}(\mathbf{k})=\hat{H}_{\mathrm{mol}}(\mathbf{k})+\hat{H}_{\mathrm{cav}}(\mathbf{k})+\hat{H}_{\mathrm{cav-mol}}(\mathbf{k}),
\end{equation}
where,
\begin{equation}
	\label{eq:Hamil}
	\begin{aligned}
		\hat{H}_{\mathrm{mol}}(\mathbf{k})=&\hbar\omega_{\mathrm{e}}\ket{+_{\mathrm{mol}}}\bra{+_{\mathrm{mol}}}+\hbar\omega_{\mathrm{e}}\ket{-_{\mathrm{mol}}}\bra{-_{\mathrm{mol}}},\\
		\hat{H}_{\mathrm{cav}}(\mathbf{k})=&\Big(E_0+\frac{\hbar^2|\mathbf{k}|^2}{2m^*}+\zeta|\mathbf{k}|\cos\phi\Big)\ket{+_{\mathrm{cav}}}\bra{+_{\mathrm{cav}}}\\
		&+\Big(E_0+\frac{\hbar^2|\mathbf{k}|^2}{2m^*}-\zeta|\mathbf{k}|\cos\phi\Big)\ket{-_{\mathrm{cav}}}\bra{-_{\mathrm{cav}}}\\
		&+\Big(-\beta_0+\beta|\mathbf{k}|^2e^{-i2\phi}\Big)\ket{+_{\mathrm{cav}}}\bra{-_{\mathrm{cav}}}\\
		&+\Big(-\beta_0+\beta|\mathbf{k}|^2e^{i2\phi}\Big)\ket{-_{\mathrm{cav}}}\bra{+_{\mathrm{cav}}},\\
		\hat{H}_{\mathrm{cav-mol}}(\mathbf{k})=&\mathbf{J}_{\mathbf{k},+}\cdot\Big(\bm{\mu}_{+}\ket{+_{\mathrm{mol}}}+\bm{\mu}_{-}\ket{-_{\mathrm{mol}}}\Big)\bra{+_{\mathrm{cav}}}\\
		&+\mathbf{J}_{\mathbf{k},-}\cdot\Big(\bm{\mu}_{+}\ket{+_{\mathrm{mol}}}+\bm{\mu}_{-}\ket{-_{\mathrm{mol}}}\Big)\bra{-_{\mathrm{cav}}}\\
		&+\mathrm{H.c.}
	\end{aligned}
\end{equation}
Here, $\mathbf{k}$ lies within the first Brillouin zone determined by the porphyrin lattice $k_x,k_y\in[-\pi/a,\pi/a]$. As we are only interested in length scales much larger than $a$, we take the continuum limit $a\to 0$ while keeping $\mu_0/a$ a constant. Therefore, terms such as the collective light-matter coupling strength, $\mathbf{J}_{\mathbf{k},\alpha}\cdot\bm{\mu}_{\alpha'}$, remain constant in this limit (see Section S1 in Supporting Information). Moreover, upon taking the continuum limit, $\hat{H}(\mathbf{k})$ does not change; only the range of $\mathbf{k}$ becomes infinitely large, $k_x,k_y\in \mathbb{R}$, that is, our system acquires complete translational invariance in the x-y plane. For such continuous systems, since $k_x,k_y\in\mathbb{R}$ is unbounded, we need to map $(k_x,k_y)$ onto a sphere which is a closed and bounded surface using stereographic projection before we compute Chern numbers \cite{silveirinha2015chern} (see Section S2 in Supporting Information).

\begin{figure*} [htpb] 
  	\includegraphics[width=\linewidth]{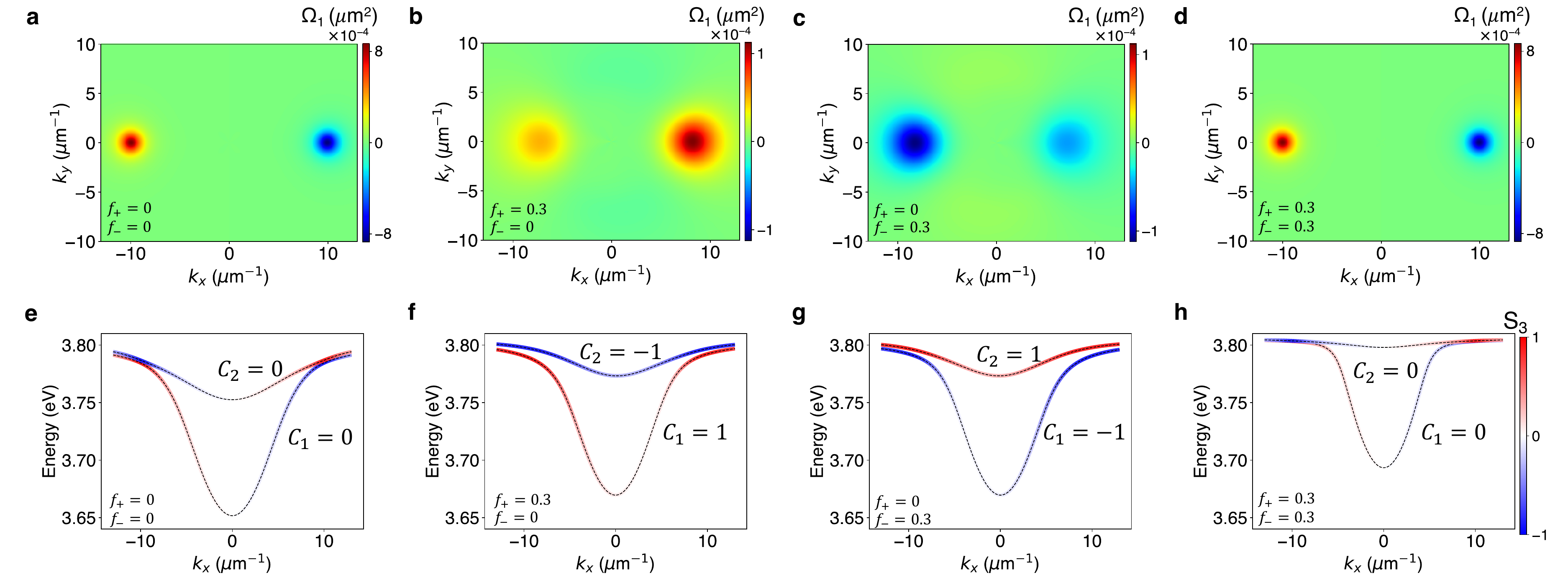}
  	\caption{\label{fig:PorphyrinBerryBand}Berry curvature and degree of circular polarization of the bands. (a-d) Berry curvature of the lowest energy band, $\Omega_{1}(\mathbf{k})$, and (e-h) a slice of the band structure at $k_y=0$ of the lower two bands, under different levels of optical pumping which create populations: (a,e) $f_+=f_-=0$, (b,f) $f_+=0.3, f_-=0$, (c,g) $f_+=0, f_-=0.3$, and (d,h) $f_+=f_-=0.3$. (e-h) The colors of the band indicate the value of the Stokes parameter, $S_3(\mathbf{k})$, which measures the degree of circular polarization of a mode (Eq. \ref{eq:S3}). The Chern numbers $C_1$ and $C_2$ of the bands are also specified and are non-zero under time-reversal symmetry (TRS) breaking, that is, when $f_+\neq f_-$. We used parameters $\beta_0 = 0.1 \mathrm{eV}$, $\beta = 9\times10^{-4} \mathrm{eV}\mu \mathrm{m}^2$, $\zeta = 2.5\times10^{-3}\mathrm{eV}\mu \mathrm{m}$, $m^* =125 \hbar^2\mathrm{eV}^{-1}\mu \mathrm{m}^{-2}$, $E_0=3.80\mathrm{eV}$ and $\hbar\omega_{\mathrm{e}}=3.81$eV (see Section S4 in Supporting Information for details).}
  \end{figure*}

When we diagonalize the Hamiltonian in Eq. \ref{eq:Hamil}, we obtain four bands which we label with $l=1,2,3,4$ in increasing order of energy. In Fig. \ref{fig:PorphyrinBerryBand}a we plot the Berry curvature, $\Omega_{1}(\mathbf{k})$, of the lowest band $l=1$, and in Fig. \ref{fig:PorphyrinBerryBand}e we plot the $k_y=0$ slice of the band structure of the two bands lowest in energy, $l=1,2$. As expected, in the absence of optical pumping, this system preserves TRS, which can be verified using the condition on Berry curvature $\Omega_{l}(\mathbf{k})=-\Omega_{l}(-\mathbf{k})$, and the Chern numbers of the all the bands $C_l=0$. Also, note that, the smallest splitting between the lower two bands within $-13 \mu$m$^{-1}<k_x,k_y<13\mu$m$^{-1}$ is $ \sim 2.8$meV which is larger than the linewidth of the transition in porphyrin at 4K ($\sim0.5$meV) \cite{even1982isolated,voelker1977homogeneous}.

\subsection*{Optical pumping}
Optical pumping can saturate the electronic transitions of a system. This leads to reduction in the effective light-matter coupling strength, and, therefore, Rabi splitting contraction \cite{daskalakis2014nonlinear, xiang2018two, yagafarov2020mechanisms}. For instance, when the pump excites a fraction of molecules, $f_{\mathrm{E}}$, to the excited state and the remaining population stays in the ground state, $f_{\mathrm{G}}$, it results in Rabi splitting contraction proportional to $\sqrt{f_{\mathrm{G}}-f_{\mathrm{E}}}=\sqrt{1-2f_{\mathrm{E}}}$ \cite{f2018theory}.

In our system, when the molecules are optically pumped, a fraction, $f_+$, of the molecules occupy the $\ket{+_{\mathrm{mol}}}$ state, another fraction, $f_-$, occupy the $\ket{-_{\mathrm{mol}}}$ state, and the remaining fraction, $f_{\mathrm{G}}$, are in the ground state $\ket{\mathrm{G}}$. The Rabi contraction corresponding to the $\ket{\mathrm{G}}$ to $\ket{+_{\mathrm{mol}}}$ transition should then be proportional to $\sqrt{f_{\mathrm{G}}-f_+}$ which equals $\sqrt{1-f_--2f_+}$  since $f_{\mathrm{G}}+f_++f_-=1$. Similarly, the contraction should be proportional to $\sqrt{1-f_+-2f_-}$ for the $\ket{\mathrm{G}}$ to $\ket{-_{\mathrm{mol}}}$ transition. This difference in light-matter coupling when $f_+\neq f_-$ effectively introduces 2D chirality into the system \cite{salij2022chiral}. 

To derive an effective Hamiltonian under optical pumping, we use Heisenberg equations of motion and make a mean-field approximation following the approach of Ribeiro \textit{et al}. \cite{f2018theory} (see Section S3 in Supporting Information). We then obtain the effective Hamiltonian,
\begin{equation}\label{eq:pumpHamil}
	\hat{H}^{\mathrm{eff}}(\mathbf{k})=\hat{H}^{\mathrm{eff}}_{\mathrm{mol}}(\mathbf{k})+\hat{H}^{\mathrm{eff}}_{\mathrm{cav}}(\mathbf{k})+\hat{H}^{\mathrm{eff}}_{\mathrm{cav-mol}}(\mathbf{k}),
\end{equation}
where,
\begin{equation}\label{eq:pumpHamilexpand}
	\begin{aligned}
		\hat{H}^{\mathrm{eff}}_{\mathrm{mol}}(\mathbf{k})=&\hbar\omega_{\mathrm{e}}\ket{+_{\mathrm{mol}}}'\bra{+_{\mathrm{mol}}}'+\hbar\omega_{\mathrm{e}}\ket{-_{\mathrm{mol}}}'\bra{-_{\mathrm{mol}}}',\\
		\hat{H}^{\mathrm{eff}}_{\mathrm{cav}}(\mathbf{k})=&\Big(E_0+\frac{\hbar^2|\mathbf{k}|^2}{2m^*}+\zeta|\mathbf{k}|\cos\phi\Big)\ket{+_{\mathrm{cav}}}'\bra{+_{\mathrm{cav}}}'\\
		&+\Big(E_0+\frac{\hbar^2|\mathbf{k}|^2}{2m^*}-\zeta|\mathbf{k}|\cos\phi\Big)\ket{-_{\mathrm{cav}}}'\bra{-_{\mathrm{cav}}}'\\
		&+\Big(-\beta_0+\beta|\mathbf{k}|^2e^{-i2\phi}\Big)\ket{+_{\mathrm{cav}}}'\bra{-_{\mathrm{cav}}}'\\
		&+\Big(-\beta_0+\beta|\mathbf{k}|^2e^{i2\phi}\Big)\ket{-_{\mathrm{cav}}}'\bra{+_{\mathrm{cav}}}',\\
		\hat{H}^{\mathrm{eff}}_{\mathrm{cav-mol}}(\mathbf{k})=&\mathbf{J}_{\mathbf{k},+}\cdot\Big(\sqrt{1-f_--2f_+}\bm{\mu}_{+}\ket{+_{\mathrm{mol}}}'\\
		&+\sqrt{1-f_+-2f_-}\bm{\mu}_{-}\ket{-_{\mathrm{mol}}}'\Big)\bra{+_{\mathrm{cav}}}'\\
		&+\mathbf{J}_{\mathbf{k},-}\cdot\Big(\sqrt{1-f_--2f_+}\bm{\mu}_{+}\ket{+_{\mathrm{mol}}}'\\
		&+\sqrt{1-f_+-2f_-}\bm{\mu}_{-}\ket{-_{\mathrm{mol}}}'\Big)\bra{-_{\mathrm{cav}}}'+\mathrm{H.c.}
	\end{aligned}
\end{equation}
Here, the states $\ket{\gamma}'$ are different from states $\ket{\gamma}$ in eq. \ref{eq:Hamil}, where $\gamma=\pm_{\mathrm{mol}},\pm_{\mathrm{cav}}$. As expected, the light-matter coupling terms are scaled by factors $\sqrt{1-f_\mp-2f_\pm}$ which is a consequence of the commutation relation in eq. \ref{eq:commutator} (see Section S3 in Supporting Information).

If the pump pulse is circularly polarized, $f_+\neq f_-$,  the Rabi contraction factor that multiplies the light-matter coupling differs for transitions to the $\ket{+_{\mathrm{mol}}}$ and $\ket{-_{\mathrm{mol}}}$ states; as a result, time-reversal symmetry is broken. Consequently, when $f_+>f_-$, we find that bands 1 and 2 have non-zero Chern numbers +1 and -1 (Fig. \ref{fig:PorphyrinBerryBand}f). Under the opposite condition, $f_+<f_-$, the Chern numbers reverse sign as seen in Fig. \ref{fig:PorphyrinBerryBand}g. When $f_+=f_-$, TRS is preserved, and all bands have Chern number 0 as seen in Fig. \ref{fig:PorphyrinBerryBand}e and \ref{fig:PorphyrinBerryBand}h. In Fig. \ref{fig:PorphyrinBerryBand}b-c, we plot the computed Berry curvature when $f_+\neq f_-$ and due to broken TRS, we find $\Omega_{l}(\mathbf{k})\neq-\Omega_{l}(-\mathbf{k})$. Non-zero values of Berry curvature are found at $k_x\sim \pm8\mu$m$^{-1}$, $k_y\sim 0\mu$m$^{-1}$ when $f_+=0.3,f_-=0$ or $f_+=0,f_-=0.3$. To measure the Berry curvature of the bands at these values of $\mathbf{k}$, the linewidths of the cavity modes and the molecular transitions need to be less than $10$meV as the energy splittings between the bands are $10-15$meV.

\begin{figure}[!t]
	\centering
	\includegraphics[width=\columnwidth]{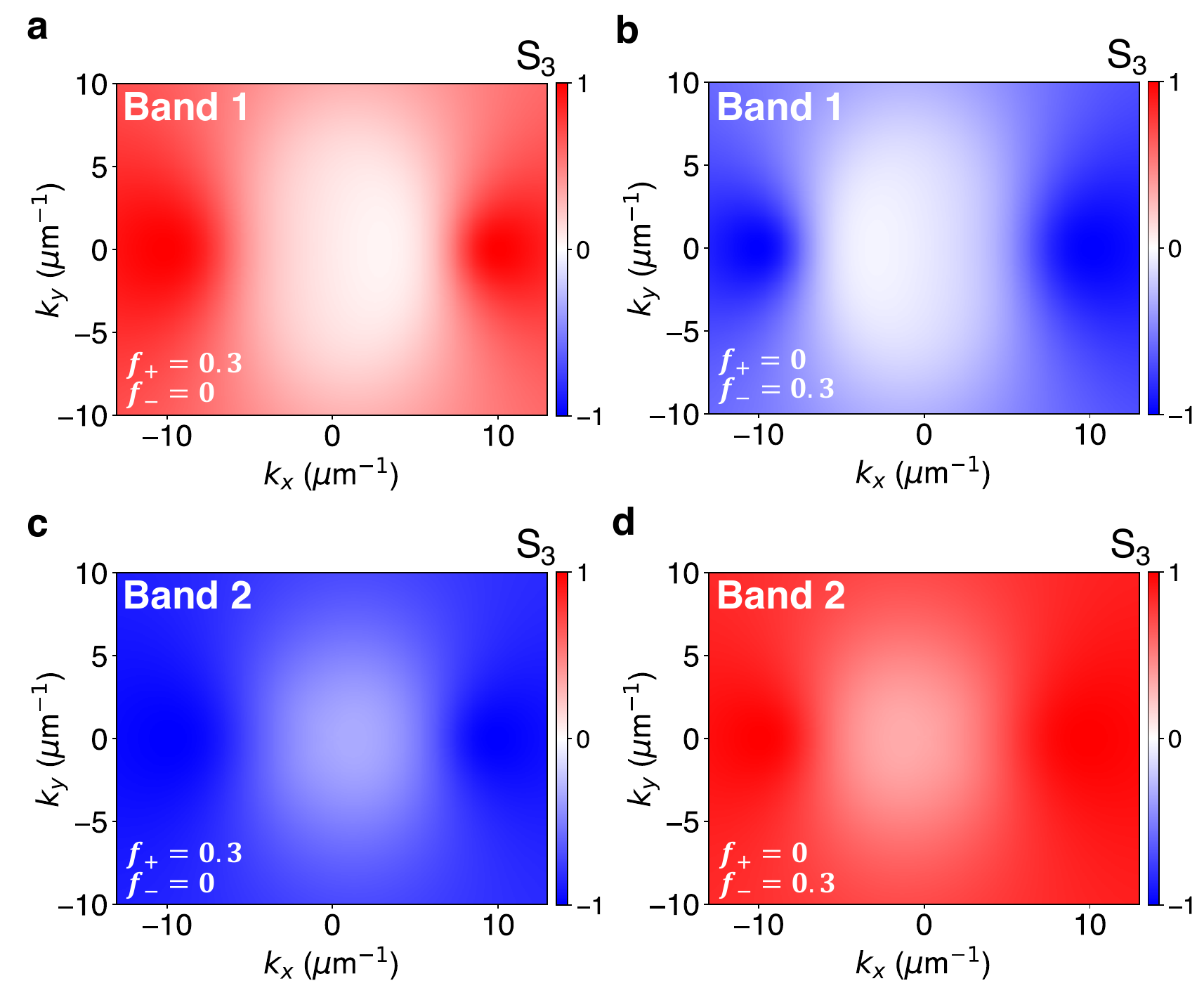}
	\caption{\label{fig:PorphyrinS3}The Stokes parameter, $S_3(\mathbf{k})$, which is a measure of the degree of circular polarization of a mode (Eq. \ref{eq:S3}), under pumping with (a,c) $\sigma_+$ polarized light which creates populations $f_+=0.3$, $f_-=0$ and (b,d) $\sigma_-$ polarized light which creates populations $f_+=0$, $f_-=0.3$ of the two lowest energy bands (Band 1 and 2 as indicated in the inset).  We used parameters $\beta_0 = 0.1 \mathrm{eV}$, $\beta = 9\times10^{-4} \mathrm{eV}\mu \mathrm{m}^2$, $\zeta = 2.5\times10^{-3}\mathrm{eV}\mu \mathrm{m}$, $m^* =125 \hbar^2\mathrm{eV}^{-1}\mu \mathrm{m}^{-2}$, $E_0=3.80\mathrm{eV}$ and $\hbar\omega_{\mathrm{e}}=3.81$eV (see Section S4 in Supporting Information for details).}
\end{figure}

We also plot the Stokes parameter, $S_3(\mathbf{k})$, for bands 1 and 2, under pumping with circularly polarized light, in Fig. \ref{fig:PorphyrinS3}. The Stokes parameter, $S_3(\mathbf{k})$, provides information on the degree of circular polarization of the photonic component of an exciton-polariton band and is calculated as
\begin{equation}\label{eq:S3}
	S_3(\mathbf{k})=\frac{|b_{+,\mathrm{cav}}(\mathbf{k})|^2-|b_{-,\mathrm{cav}}(\mathbf{k})|^2}{|b_{+,\mathrm{cav}}(\mathbf{k})|^2+|b_{-,\mathrm{cav}}(\mathbf{k})|^2}
\end{equation}
where the eigenvectors of the band are $\ket{u_{l,\mathbf{k}}}=b_{+,\mathrm{cav}}(\mathbf{k})\ket{+_{\mathrm{cav}}}+b_{-,\mathrm{cav}}(\mathbf{k})\ket{-_{\mathrm{cav}}}+b_{+,\mathrm{mol}}(\mathbf{k})\ket{+_{\mathrm{mol}}}+b_{-,\mathrm{mol}}(\mathbf{k})\ket{-_{\mathrm{mol}}}$. In the absence of pumping, we find that within a band, one half of the modes are predominantly $\sigma_+$ polarized and the other half are $\sigma_-$ polarized (Fig. \ref{fig:PorphyrinBerryBand}e). Upon pumping with circularly polarized light, a large number of modes within each band gradually become of the same polarization as $|f_+-f_-|$ is increased (Fig. \ref{fig:PorphyrinBerryBand}f-g, Fig. \ref{fig:PorphyrinS3} and Fig. S2).

In experiments, the Berry curvature of photon bands in a Fabry-Perot cavity can be extracted from the components of the Stokes vector \cite{ren2021nontrivial,lempicka2022electrically}. In the case of exciton-polariton bands, the Berry curvature can be measured experimentally using the Stokes vector when the bands of the system can be separated into pairs of bands that are effectively described by separate $2\cross2$ Hamiltonians. At each $\mathbf{k}$, the Stokes vector can describe a state in a two-dimensional Hilbert space; however, the Stokes vector does not contain enough information to fully specify a state in a Hilbert space of dimensions larger than two. Therefore, in our four band model, the Berry curvature (Fig. \ref{fig:PorphyrinBerryBand}a-d) can be experimentally measured by pump-probe spectroscopy only when the splitting induced by the light-matter coupling is much larger than that induced by the coupling between cavity modes because then the four polariton bands can be separated into two pairs of bands that are effectively described by separate $2\cross2$ Hamiltonians as in \cite{polimeno2021tuning}. This measurement should be feasible as long as the time delay between the pump and probe pulses is shorter than the time the system takes to depolarize and reach a state with $f_+=f_-$. The system's depolarization time depends only upon the bare molecular depolarization rate. As the depolarization timescale for porphyrins ranges from 210 fs to 1.6 ps, this measurement should be viable \cite{galli1993direct}. 

Population imbalances in the molecule or solid-state system can potentially be experimentally created in a variety of ways. One possibility is to directly excite higher energy material transitions with circularly polarized light that are within the transparency window of the cavity typically known as ``non-resonant" pumping \cite{daskalakis2014nonlinear,pickup2018optical}. If decay from those higher energy transitions into the relevant excited states happens before depolarization ensues, we will have obtained the desired population imbalances. Another possibility that bypasses the need of incoherent processes is a stimulated electronic Raman scattering with circularly polarized fields, although this scenario might require X-rays \cite{youngroadmap2018,amentresonant2011}. Finally, the population imbalance may also be created by pumping resonantly with a circularly polarized laser at $|\mathbf{k}|=\sqrt{\beta_0/\beta}$, $\phi=0$. At this angle, the coupling between the circularly polarized cavity modes is zero. Additionally, $|\mathbf{J}_{\mathbf{k},+}\cdot\bm{\mu}_{-}|\gg|\mathbf{J}_{\mathbf{k},+}\cdot\bm{\mu}_{+}|$ and $|\mathbf{J}_{\mathbf{k},-}\cdot\bm{\mu}_{+}|\gg|\mathbf{J}_{\mathbf{k},-}\cdot\bm{\mu}_{-}|$ for all $|\mathbf{k}|\ll n_z\pi/L_z$. Therefore, when the polariton mode at this $\mathbf{k}$ is pumped with circularly polarized light, the cavity mode of only the corresponding circular polarization is excited and population is transferred largely to only one of the circularly polarized molecular states. After dephasing into the molecular states (but not depolarization of the latter), the populations of the molecular states would be unequal $f_+\neq f_-$.

As the Chern numbers of bands 1 and 2 are modified through pumping with circularly polarized light, if we perform a calculation where a region of the system is pumped with $\sigma_+$ polarized light ($f_+\neq 0$ and $f_-=0$) and an adjacent region is pumped with $\sigma_-$ polarized light ($f_+= 0$ and $f_-\neq0$), we expect edge states at the boundary between these regions. However, as our Hamiltonian does not contain couplings between neighboring molecules, and the position of a molecule does not enter the Hamiltonian anywhere except through the phase of the light-matter coupling $e^{i\mathbf{k\cdot r_m}}$, the standard bulk-boundary correspondence is no longer applicable and we do not observe edge states. We do not include plots for these calculations in this work and leave it an open question whether there is an analogous statement for bulk-boundary correspondence in these types of systems. On the other hand, for exciton-polariton systems where nearest-neighbor couplings are present, edge states have been predicted and observed \cite{klembt2018exciton,karzig2015topological}.

\begin{figure*} [htpb]
	\includegraphics[width=\linewidth]{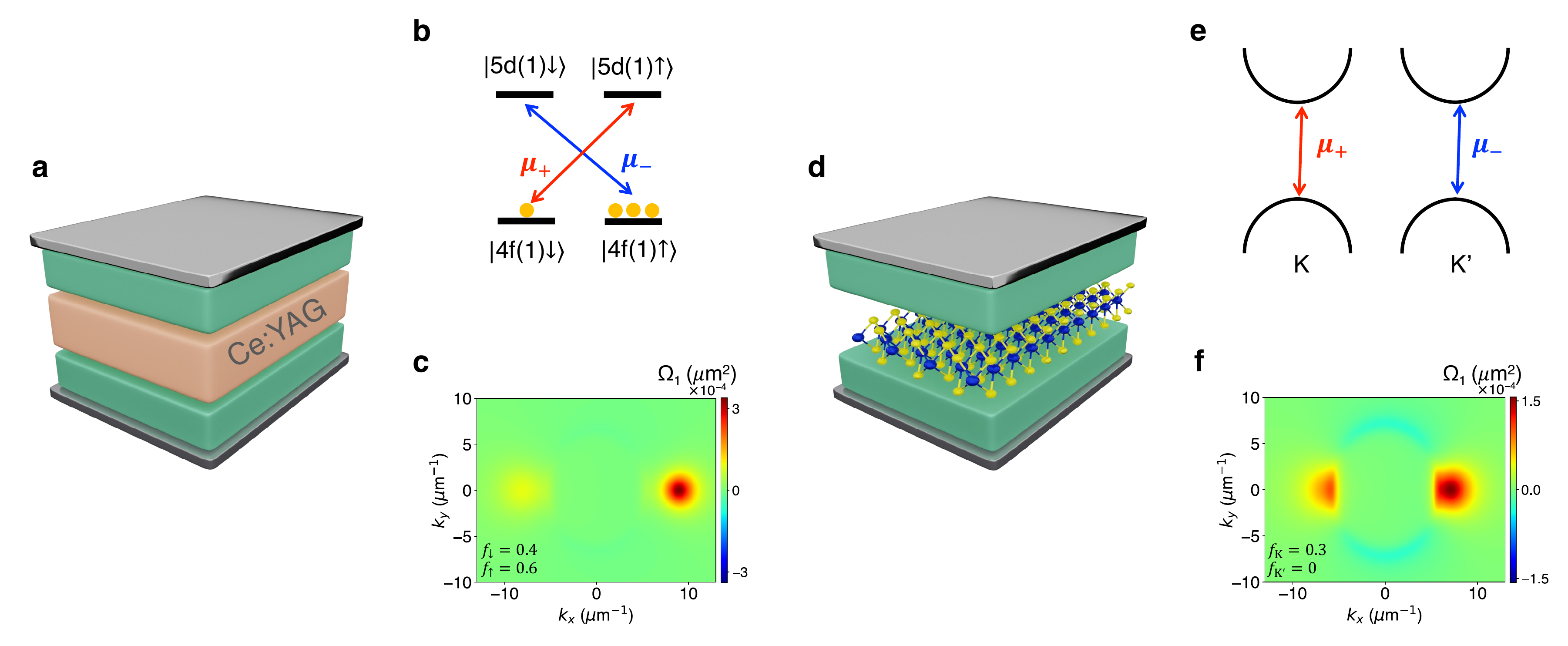}
	\caption{\label{fig:othersystems}Solid-state polariton systems where population imbalance induces non-trivial topology. (a) Illustration of Ce:YAG (salmon block) and perylene (green blocks) within a Fabry-Perot cavity. (b) Atomic levels of Ce$^{3+}$ ions embedded in Yttrium Aluminum garnet (YAG) where the yellow circles indicate the fraction $f_{\downarrow}$ of Ce$^{3+}$ ions in the $\ket{4f(1)\downarrow}$ state and the fraction $f_{\uparrow}$ in the $\ket{4f(1)\uparrow}$ state after optical pumping. The transition dipoles $\bm{\mu}_{\pm}=\mu_0(\mathbf{\hat{x}}\pm i\mathbf{\hat{y}})/\sqrt{2}$ are also indicated. (c) Berry curvature of the lowest energy band, $\Omega_{1}(\mathbf{k})$, under pumping with circularly polarized which creates populations $f_{\downarrow}=0.4$ and $f_{\uparrow}=0.6$. (d) Illustration of monolayer MoS$_2$ and perylene (green blocks) within a Fabry-Perot cavity. (e) Illustration of A-excitons in the K and K' valleys of monolayer MoS$_2$. (f) Berry curvature of the lowest energy band, $\Omega_{1}(\mathbf{k})$, under pumping with circularly polarized which creates exciton populations $f_{\mathrm{K}}=0.3$ and $f_{\mathrm
			{K'}}=0$.  We used parameters $\beta_0 = 0.1 \mathrm{eV}$, $\beta = 9\times10^{-4} \mathrm{eV}\mu \mathrm{m}^2$, $\zeta = 2.5\times10^{-3}\mathrm{eV}\mu \mathrm{m}$, $m^* =125 \hbar^2\mathrm{eV}^{-1}\mu \mathrm{m}^{-2}$, (c) $E_0=2.50 \mathrm{eV}$, $\hbar\omega_{\mathrm{e}}=2.53$eV and (f) $E_0=1.80 \mathrm{eV}$, $\hbar\omega_{\mathrm{e}}=1.855$eV (see Section S4 in Supporting Information for details). }
\end{figure*}

\subsection*{Other systems}

To emphasize that our scheme of saturating electronic transitions with circularly polarized light to modify topological properties is not limited to organic exciton-polariton systems, we compute the Berry curvature of two other polariton systems where porphyrin is replaced with (i) Ce:YAG and (ii) MoS$_2$ (Fig. \ref{fig:othersystems}a and \ref{fig:othersystems}d). Other materials can also be used in place of porphyrins, as long as they have transitions that can be selectively excited with circularly polarized light and these transitions have large enough transition dipole moments that they can couple strongly to the photon modes of a cavity. 

In Yttrium Aluminum garnet (YAG) doped with Cerium, Ce$^{3+}$ ions replace some Y$^{3+}$ and Ce$^{3+}$ has transitions that can be selectively excited with circularly polarized light. Here, each Ce$^{3+}$ has two possible ground states, one with the electron in spin up $\ket{4f(1)\uparrow}$, and the other with it in spin down $\ket{4f(1)\downarrow}$. Similarly, it has a degenerate pair of excited spin states $\ket{5d(1)\uparrow}$ and $\ket{5d(1)\downarrow}$. The $\ket{4f(1)\downarrow}\leftrightarrow \ket{5d(1)\uparrow}$ transition has $\sim400$ times larger oscillator strength for excitation with $\sigma_+$ polarized light than with $\sigma_-$ polarized light, therefore, we take the transition dipole moment to be $\bm{\mu}_+$ (Fig. \ref{fig:othersystems}b) \cite{kolesov2013mapping}. Similarly, we take the transition dipole to be $\bm{\mu}_-$ for the $\ket{4f(1)\uparrow}\leftrightarrow \ket{5d(1)\downarrow}$ transition (Fig. \ref{fig:othersystems}b). The transitions in Ce:YAG do couple to photon modes, however, to the best of our knowledge, strong coupling has not been reported in the literature \cite{moerland2018nanoscale,rodriguez2012light}. Nevertheless, strong light-matter coupling has been achieved with a similar system: Nd$^{3+}$ doped YSO and YVO crystals \cite{zhong2015nanophotonic,zhong2017interfacing}, and based on our calculations, with a 0.1$\mu$m thick sample of Ce:YAG at concentration 1\% Ce$^{3+}$ (relative to Y$^{3+}$), we should be able to attain strong coupling with photon modes in a Fabry-Perot cavity (see Section S4 in Supporting Information). 

Under thermal equilibrium, the populations of the $\ket{4f(1)\uparrow}$ and $\ket{4f(1)\downarrow}$ states are equal. However, under pumping with pulses of $\sigma_+$ polarization, in the presence of a small magnetic field $\sim0.049$T, the population of $\ket{4f(1)\uparrow}$ will exceed that of $\ket{4f(1)\downarrow}$ because population is selectively removed from $\ket{4f(1)\downarrow}$ and added to $\ket{5d(1)\uparrow}$ by the circularly polarized pulses, but decay from the excited $\ket{5d(1)\uparrow}$ state to the two ground states has equal probability \cite{siyushev2014coherent}. In principle, a magnetic field is not required; however, as we do not know the spin relaxation time in the absence of the magnetic field, we report the magnetic field used in the experimental study \cite{siyushev2014coherent}. Under optical pumping with circularly polarized light, the $5d$ states will have very small populations which we take to be zero, while the $\ket{4f(1)\downarrow}$ and $\ket{4f(1)\uparrow}$ states will have unequal populations $f_{\downarrow}$ and $f_{\uparrow}$, respectively; here, $f_{\downarrow}+f_{\uparrow}=1$. Optically pumped Ce:YAG can then be modeled using the effective Hamiltonian in eq. \ref{eq:pumpHamil} and \ref{eq:pumpHamilexpand}, with $\ket{\pm_{\mathrm{mol}}}'\to \ket{5d(1)\uparrow/\downarrow}$ and $\sqrt{1-f_\mp-2f_\pm}\to \sqrt{f_{\downarrow/\uparrow}}$. The large spin relaxation time of $\sim4.5$ ms makes this system particularly well-suited for our scheme because it maintains $f_{\downarrow}\neq f_{\uparrow}$, and hence non-zero Chern invariants, for an extended period of time \cite{siyushev2014coherent}. In Fig. \ref{fig:othersystems}c we plot Berry curvature of the lowest band of a perylene filled cavity strongly coupled with Ce:YAG, where $f_{\downarrow}=0.4$ and $f_{\uparrow}=0.6$ (see Section S4 in Supporting Information for values of other parameters).

TMDs, such as single-layer MoS$_2$, display optically controllable valley polarization and could also be used in place of porphyrins \cite{mak2012control,zeng2012valley,pattanayak2022probing}. Due to lack of inversion symmetry in these systems, the K and K' valleys are inequivalent; this results in optical selection rules that allow selective creation of excitons at K and K' valleys with $\sigma_+$ and $\sigma_-$ polarized light, respectively \cite{sun2019separation,peng2022twisted}. Additionally, strong light-matter coupling has been observed when monolayer MoS$_2$ is placed within a Fabry-Perot cavity \cite{liu2015strong,hu2020recent}. This system has depolarization times of $\sim200$fs - $5$ps making it possible to measure Berry curvature using pump-probe spectroscopy before depolarization occurs \cite{dal2015ultrafast,chen2017valley}. We model this exciton-polariton system (Fig. \ref{fig:othersystems}d) using eq. \ref{eq:pumpHamil} and eq. \ref{eq:pumpHamilexpand} (we focus on the A-exciton, see Section S4 in Supporting Information for parameters) with $\ket{+_{\mathrm{mol}}}\to \ket{\mathrm{K}}$, $\ket{-_{\mathrm{mol}}}\to \ket{\mathrm{K}'}$ and $\sqrt{1-f_{\mp}-2f_{\pm}}\to\sqrt{1-2f_{\mathrm{K}/\mathrm{K}'}}$. In Fig. \ref{fig:othersystems}f we plot the Berry curvature of the lowest band when $f_{\mathrm{K}}=0.3$ and $f_{\mathrm{K}'}=0$. Unfortunately, significant Rabi contraction upon optical pumping has not been experimentally observed in these systems which will make it challenging to observe Berry curvature as in Fig. \ref{fig:othersystems}f since our model relies on saturation effects. However, for exciton polaritons formed from monolayer TMDs, even if Rabi contraction through resonant optical pumping may not produce the intended effect, off-resonant optical pumping can break the degeneracy of excitons in the K and K' valleys through optical stark effect \cite{lamountain2021valley}, and this may have interesting consequences for the Berry curvature. Additionally, if bilayer MoS$_2$ is used in place of monolayer MoS$_2$, effects on the Berry curvature described in our work may be more pronounced as bilayer MoS$_2$ hosts interlayer excitons which possess large optical nonlinearities; specifically, they display saturation and Rabi contraction under strong coupling \cite{datta2022highly,louca2022nonlinear}.

Finally, so far we have only considered replacing porphyrin with a different material, such as MoS$_2$ or Ce:YAG. In addition to this, perylene can also be replaced with other suitable materials. In our work, we choose to use a cavity filled with perylene because we do not want degeneracy at any $\mathbf{k}$ within the photon bands. Other systems also satisfy this requirement and could be used instead. For instance, we could use an electrically tunable, highly anisotropic, liquid-crystal cavity with well separated H and V polarized photon modes \cite{rechcinska2019engineering,muszynski2022realizing}. A perovskite cavity is another potential candidate due to its high anisotropy, and optical pumping may help lift the degeneracy of polariton modes in this system \cite{polimeno2021tuning}. Additionally, other photonic structures can also be used instead of a cavity, as long as the photon bands are not degenerate at any $\mathbf{k}$ and have non-zero light-matter coupling at all $\mathbf{k}$.

In our analysis, we have disregarded the explicit role of vibrational modes, which is a reasonable assumption for rigid molecular systems (such as porphyrins and phthalocyanines \cite{renger2002on}) and solid-state systems as their electron-phonon (vibronic) couplings tend to be small.

\section*{Conclusion}
In summary, we show that TRS can be broken in organic exciton-polariton systems through selectively saturating electronic transitions with a circularly polarized pump and that the resulting bands possess non-zero Chern invariants. In particular, we demonstrate this theoretically for a Fabry-Perot cavity filled with porphyrin and perylene. The Berry curvature of the more photonic parts of the bands of this system can be measured experimentally using pump-probe spectroscopy, as long as the time delay is shorter than the depolarization time for porphyrin (210fs-1.6ps) \cite{galli1993direct}, and this will reveal non-zero Berry curvature and Chern number under circularly polarized pumping. Our scheme relies on Rabi contraction from saturation of optical transitions. It is important to note that edge states do not emerge in our system despite non-zero Chern invariants as our model does not contain sufficient positional information about the molecules or the unit cells. Bleu \textit{et al}. \cite{bleu2017photonic} have previously proposed breaking TRS in inorganic exciton-polariton systems through pumping with circularly polarized light, however, their work relies on polariton condensation and having patterned lattices. Finally, we demonstrate that saturating electronic transitions to modify topology is not limited to organic systems. To illustrate this, we calculate the Berry curvature and Chern numbers of exciton-polariton bands of two other systems under optical pumping: (a) Ce:YAG and (b) monolayer MoS$_2$, and find similar results as the organic exciton-polariton case. In view of recent developments on electrically tuning the Berry curvature of liquid-crystal and perovskite filled cavities \cite{rechcinska2019engineering,lempicka2022electrically}, our work provides an additional control knob to optically tune the Berry curvature of exciton-polariton systems using circularly polarized light. Additionally, ultrafast control of topological properties of systems with light may find use in nonreciprocal and nonlinear optoelectronic devices.

\subsection*{Code availability}
Data underlying the results presented
in this paper may be obtained from the corresponding author upon request. They were generated using code available at \textcolor{blue}{https://github.com/SindhanaPS/Topological\_Polaritons\_Submission}.

\begin{acknowledgements}
S.P.-S. acknowledges support from NSF Grant No. CAREER CHE 1654732 for the development of the model and calculations. The conceptualization of the molecular and solid-state systems was guided by N.P.S. and J.Y.-Z. as part of the Center for Molecular Quantum Transduction (CMQT), an Energy Frontier Research Center funded by the U.S. Department of Energy, Office of Science, Basic Energy Sciences under Award No. DE-SC0021314. S.P.-S. thanks Kai Schwennicke and Stephan van den Wildenberg for useful discussions.
\end{acknowledgements}

\end{document}


\title{Supplementary information: Molecular and solid-state topological polaritons induced by population imbalance}

\author{Sindhana Pannir-Sivajothi}
\affiliation{Department of Chemistry and Biochemistry, University of California San Diego, La Jolla, California 92093, USA}
\author{Nathaniel P. Stern}
\affiliation{Department of Physics and Astronomy, Northwestern University, Evanston, Illinois 60208, USA}
\author{Joel Yuen-Zhou}
\email{joelyuen@ucsd.edu}
\affiliation{Department of Chemistry and Biochemistry, University of California San Diego, La Jolla, California 92093, USA}
	
	\maketitle
	\section{Light-matter coupling}
	The light-matter coupling part of the total Hamiltonian under the electric dipole approximation is,
	\begin{equation}
		\begin{aligned}
			\hat{H}_{\mathrm{cav-mol}}=&\sum_{\mathbf{m}}\sum_{\mathbf{k},\alpha}-\bm{\hat{\mu}}_{\mathbf{m}}\cdot\mathbf{\hat{E}}_{\mathbf{k},\alpha}(\mathbf{r}_{\mathbf{m}},0),\\
			=&\sum_{\mathbf{m}}\sum_{\mathbf{k},\alpha}-\Big[\sum_{\alpha'=\pm}(\bm{\mu}_{\alpha'}\hat{\sigma}_{\mathbf{m},\alpha'}^{\dagger}+\bm{\mu}_{\alpha'}^*\hat{\sigma}_{\mathbf{m},\alpha'})\Big]\cdot\mathbf{\hat{E}}_{\mathbf{k},\alpha}(\mathbf{r}_{\mathbf{m}},0),
		\end{aligned}
	\end{equation}
	where $\bm{\mu}_{\alpha'}=\bm{\mu}_{\mathbf{m},\alpha'}=\bra{\mathbf{m},\alpha'_{\mathrm{mol}}}\bm{\hat{\mu}}\ket{\mathbf{m},\mathrm{G}}$ is independent of $\mathbf{m}$ since we assume that all porphyrin molecules lie flat in the x-y plane and are oriented. The electric field operator of the mode labeled by $\mathbf{k}$ and $\alpha$ is 
	\begin{equation}
		\mathbf{\hat{E}}_{\mathbf{k},\alpha}(\mathbf{r},z)=\sqrt{\frac{\hbar\omega_{\mathbf{k},\alpha}}{2V\varepsilon\epsilon_0}}\Big(\mathbf{f}^*_{\mathbf{k},\alpha}(\mathbf{r},z)\hat{a}^{\dagger}_{\mathbf{k},\alpha}+\mathbf{f}_{\mathbf{k},\alpha}(\mathbf{r},z)\hat{a}_{\mathbf{k},\alpha}\Big).
	\end{equation}
	Here, $V=L_xL_yL_z$ is the volume of the box we consider, where as mentioned in the main manuscript, we apply periodic boundary conditions along the $x$ and $y$ directions. From here on, we will call the in-plane area of the box $A=L_xL_y$.
	Here, $\mathbf{f}_{\mathbf{k},\alpha}(\mathbf{r},z)$ is the mode profile and it satisfies \cite{fabre2020modes}
	\begin{equation}
		\int d\mathbf{r}\int_{0}^{L_z}dz \mathbf{f}_{\mathbf{k},\alpha}^{*}(\mathbf{r},z)\mathbf{f}_{\mathbf{k},\alpha}(\mathbf{r},z)=L_zA.
	\end{equation}
	For the TE and TM modes \cite{zoubi2005microscopic}, 
	\begin{equation}\label{eq:electric}
		\begin{aligned}
			\mathbf{f}_{\mathbf{k},\mathrm{TE}}(\mathbf{r},z)=&e^{i\mathbf{k}\cdot\mathbf{r}}\sqrt{2}\sin\Bigg[\frac{n_z\pi}{L_{z}} \Big(z+\frac{L_{z}}{2}\Big)\Bigg]\bm{\hat{\phi}},\\
			\mathbf{f}_{\mathbf{k},\mathrm{TM}}(\mathbf{r},z)=&e^{i\mathbf{k}\cdot\mathbf{r}}\sqrt{\frac{2}{|\mathbf{k}|^2+\big(\frac{n_z\pi}{L_{z}}\big)^2}}\Bigg\{\Big(\frac{n_z\pi}{L_{z}}\Big)\sin\Bigg[\frac{n_z\pi}{L_{z}}\Big(z+\frac{L_{z}}{2}\Big)\Bigg]\bm{\hat{\rho}}-i|\mathbf{k}|\cos\Bigg[\frac{n_z\pi}{L_{z}}\Big(z+\frac{L_{z}}{2}\Big)\Bigg]\mathbf{\hat{z}}\Bigg\}.
		\end{aligned}
	\end{equation}
	We make the rotating-wave approximation,
	\begin{equation}\label{eq:rotatwave}
		\begin{aligned}
			\hat{H}_{\mathrm{cav-mol}}
			=&\sum_{\mathbf{m}}\sum_{\mathbf{k},\alpha}-\Big[\sum_{\alpha'=\pm}(\bm{\mu}_{\alpha'}\hat{\sigma}_{\mathbf{m},\alpha'}^{\dagger}+\bm{\mu}_{\alpha'}^*\hat{\sigma}_{\mathbf{m},\alpha'})\Big]\cdot\Big[\sqrt{\frac{\hbar\omega_{\mathbf{k},\alpha}}{2V\varepsilon\epsilon_0}}\Big(\mathbf{f}_{\mathbf{k},\alpha}^*(\mathbf{r}_{\mathbf{m}},0)\hat{a}^{\dagger}_{\mathbf{k},\alpha}+\mathbf{f}_{\mathbf{k},\alpha}(\mathbf{r}_{\mathbf{m}},0)\hat{a}_{\mathbf{k},\alpha}\Big)\Big],\\
			\approx&\sum_{\mathbf{m},\alpha'}\sum_{\mathbf{k},\alpha}-\sqrt{\frac{\hbar\omega_{\mathbf{k},\alpha}}{2V\varepsilon\epsilon_0}}\Big[\bm{\mu}_{\alpha'}\cdot\mathbf{f}_{\mathbf{k},\alpha}(\mathbf{r}_{\mathbf{m}},0)\hat{\sigma}_{\mathbf{m},\alpha'}^{\dagger}\hat{a}_{\mathbf{k},\alpha}+\bm{\mu}_{\alpha'}^*\cdot\mathbf{f}_{\mathbf{k},\alpha}^*(\mathbf{r}_{\mathbf{m}},0)\hat{\sigma}_{\mathbf{m},\alpha'}\hat{a}^{\dagger}_{\mathbf{k},\alpha}\Big],\\
			=&\sum_{\mathbf{m},\alpha'}\sum_{\mathbf{k},\alpha}\Big[\frac{e^{i\mathbf{k\cdot r_m}}}{\sqrt{N_xN_y}}(\bm{\mu}_{\alpha'}\cdot\mathbf{J}_{\mathbf{k},\alpha})\hat{\sigma}_{\mathbf{m},\alpha'}^{\dagger}\hat{a}_{\mathbf{k},\alpha}+\frac{e^{-i\mathbf{k\cdot r_m}}}{\sqrt{N_xN_y}}(\bm{\mu}_{\alpha'}^*\cdot\mathbf{J}_{\mathbf{k},\alpha}^*)\hat{\sigma}_{\mathbf{m},\alpha'}\hat{a}^{\dagger}_{\mathbf{k},\alpha}\Big],
		\end{aligned}
	\end{equation}
	where $\mathbf{J}_{\mathbf{k},\alpha}=-\sqrt{N_xN_y}\sqrt{\frac{\hbar\omega_{\mathbf{k},\alpha}}{2V\varepsilon\epsilon_0}}e^{-i\mathbf{k}\cdot\mathbf{r}}\mathbf{f}_{\mathbf{k},\alpha}(\mathbf{r},0)$ and $\bm{\mu}_{\alpha'}\cdot\mathbf{J}_{\mathbf{k},\alpha}$ is the collective light-matter coupling strength.
	
	The annihilation operators of photon modes polarized along the horizontal (H) or x-axis and vertical (V) or y-axis are $\hat{a}_{\mathbf{k},\mathrm{H}}$ and $\hat{a}_{\mathbf{k},\mathrm{V}}$, respectively. They are related to $\alpha=\pm$ polarized modes through $\hat{a}_{\mathbf{k},\pm}=\frac{1}{\sqrt{2}}(\hat{a}_{\mathbf{k},\mathrm{H}} \mp i\hat{a}_{\mathbf{k},\mathrm{V}})$ \cite{martinelli2017polarization}. In addition, we assume that they are related to the TM and TE modes through $\hat{a}_{\mathbf{k},\mathrm{TM}}=\cos\phi\hat{a}_{\mathbf{k},\mathrm{H}}+\sin\phi\hat{a}_{\mathbf{k},\mathrm{V}}$ and $\hat{a}_{\mathbf{k},\mathrm{TE}}=-\sin\phi\hat{a}_{\mathbf{k},\mathrm{H}}+\cos\phi\hat{a}_{\mathbf{k},\mathrm{V}}$. Using this, we obtain the relationship between $\hat{a}_{\mathbf{k},\mathrm{TE}}$, $\hat{a}_{\mathbf{k},\mathrm{TM}}$ and $\hat{a}_{\mathbf{k},+}$, $\hat{a}_{\mathbf{k},-}$ modes to be,
	\begin{equation}
		\begin{aligned}
			\hat{a}_{\mathbf{k},\mathrm{TM}}=&\frac{1}{\sqrt{2}}\Big(e^{i\phi}\hat{a}_{\mathbf{k},+}+e^{-i\phi}\hat{a}_{\mathbf{k},-}\Big),\\
			\hat{a}_{\mathbf{k},\mathrm{TE}}=&\frac{1}{\sqrt{2}}\Big(ie^{i\phi}\hat{a}_{\mathbf{k},+}-ie^{-i\phi}\hat{a}_{\mathbf{k},-}\Big).
		\end{aligned}
	\end{equation}
	It is important to note that, based on these relationships and \ref{eq:electric}, the $\alpha=$H/V modes are not completely linearly polarized and the $\alpha=\pm$ modes are not completely circularly polarized when $|\mathbf{k}|$ becomes comparable with $n_z\pi/L_z$. We also find,
	\begin{equation}
		\begin{aligned}
			\mathbf{J}_{\mathbf{k},+}=&\frac{e^{i\phi}}{\sqrt{2}}\Big(\mathbf{J}_{\mathbf{k},\mathrm{TM}}+i\mathbf{J}_{\mathbf{k},\mathrm{TE}}\Big),\\
			\mathbf{J}_{\mathbf{k},-}=&\frac{e^{-i\phi}}{\sqrt{2}}\Big(\mathbf{J}_{\mathbf{k},\mathrm{TM}}-i\mathbf{J}_{\mathbf{k},\mathrm{TE}}\Big).
		\end{aligned}
	\end{equation}

	To keep the collective coupling strength $\bm{\mu}_{\alpha'}\cdot\mathbf{J}_{\mathbf{k},\alpha}$ constant while taking the $a\to 0$ limit, we take the magnitude of the collective transition dipole of the bright state $\sqrt{N_xN_y}\mu_0$ over square root of the quantization area of the photon mode $\sqrt{A}$ to be a constant; that is, we keep  $\sqrt{\rho_A}\mu_0=\mu_0/a$ a constant, where $\rho_A=N_xN_y/A$ is the areal density of quantum emitters.
	\begin{equation}
		\begin{aligned}
			\mathbf{J}_{\mathbf{k},\alpha}=&-\sqrt{\rho_A}\sqrt{\frac{\hbar\omega_{\mathbf{k},\alpha}}{2L_z\varepsilon\epsilon_0}}e^{-i\mathbf{k}.\mathbf{r}}\mathbf{f}_{\mathbf{k},\alpha}(\mathbf{r},0)\\
			=&-\frac{1}{a}\sqrt{\frac{\hbar\omega_{\mathbf{k},\alpha}}{2L_z\varepsilon\epsilon_0}}e^{-i\mathbf{k}.\mathbf{r}}\mathbf{f}_{\mathbf{k},\alpha}(\mathbf{r},0).
		\end{aligned}
	\end{equation}
	
	\section{Chern number calculation}
	\begin{figure*} [htpb] 
		\includegraphics[width=\linewidth]{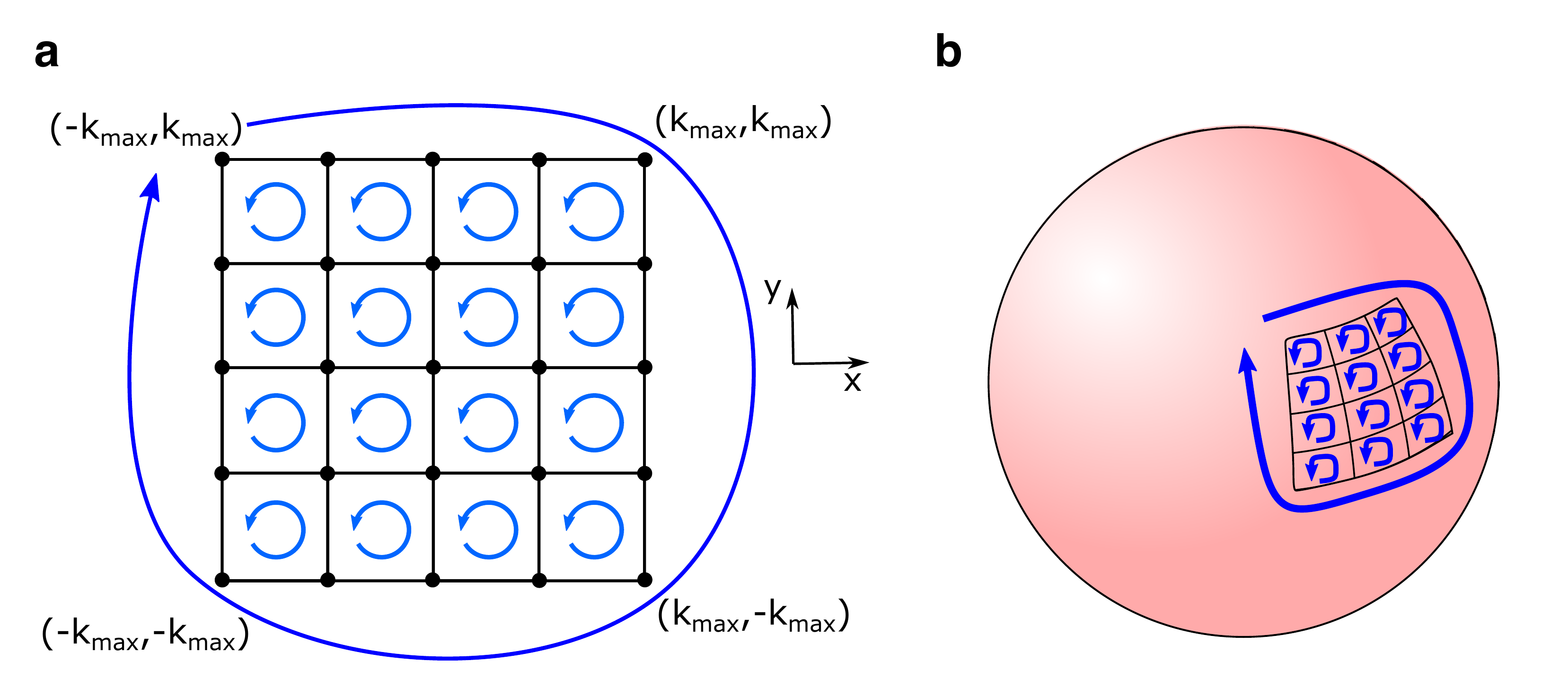}
		\caption{(a) This is a cartoon figure that demonstrates the way Berry flux and Chern number are computed in our system. The small squares are the plaquettes over which Berry flux is computed. The blue arrows specify the orientation used for Berry flux computation. Note that the direction is opposite for the small squares and the large square. (b) Same as (a), but placed on a sphere. Here, it is more clear that the direction of the arrow for the large square indicates the way Berry flux is computed for the giant plaquette covering the rest of the sphere.}
	\end{figure*}

	For the Chern invariant to be an integer, it is important that the Berry curvature is integrated over a closed and bounded surface \cite{asboth2016short}. For periodic systems with a finite period, the Brillouin zone is a torus which satisfies this requirement. However, for a continuous system, $(k_x,k_y)$ lies on an unbounded plane; for such systems, Silveirinha \cite{silveirinha2015chern} proposed mapping this infinitely large plane onto a sphere to compute the Chern number. This is the procedure we follow in our work. We discretize k-space and compute the Berry flux in each plaquette within a square-shaped region in k-space, $-k_{\mathrm{max}}\le k_x,k_y\le k_{\mathrm{max}}$ \cite{fukui2005chern,asboth2016short} (Fig. S1a and S1b). The entire region that satisfies the condition $k_x,k_y>k_{\mathrm{max}}$ or $k_x,k_y<-k_{\mathrm{max}}$ is taken as a single giant plaquette (Fig. S1b), and the Berry flux within this region is computed by taking the Berry phase along the boundary of the plaquette but in a direction opposite to that used to compute Berry flux for plaquettes within the square $-k_{\mathrm{max}}\le k_x,k_y\le k_{\mathrm{max}}$ as indicated in Fig. S1a and S1b. To ensure that we obtain a converged Chern number, we calculate the Chern number for different $k_{\mathrm{max}}$ and find that, for our system, once $k_{\mathrm{max}} \gtrsim 100 \mu$m$^{-1}$, the Chern number converges to $C_1=\pm 1, C_2=\mp 1, C_3=0,$ and $C_4=0$ when $f_+\neq f_-$ with $|f_+-f_-|\gtrsim 0.11$. Smaller differences between $f_+$ and $f_-$, $|f_+-f_-|\lesssim 0.11$ require larger $k_{\mathrm{max}}$ for convergence. This is not a problem for the $f_+=f_-$ case because the Chern invariant will always be zero due to time-reversal symmetry $\Omega_l(\mathbf{k})=-\Omega_l(-\mathbf{k})$, and we can use $k_{\mathrm{max}}\approx 100\mu$m$^{-1}$ to compute it.

	\section{Optical pumping}
	\begin{figure}
		\includegraphics[width=\textwidth]{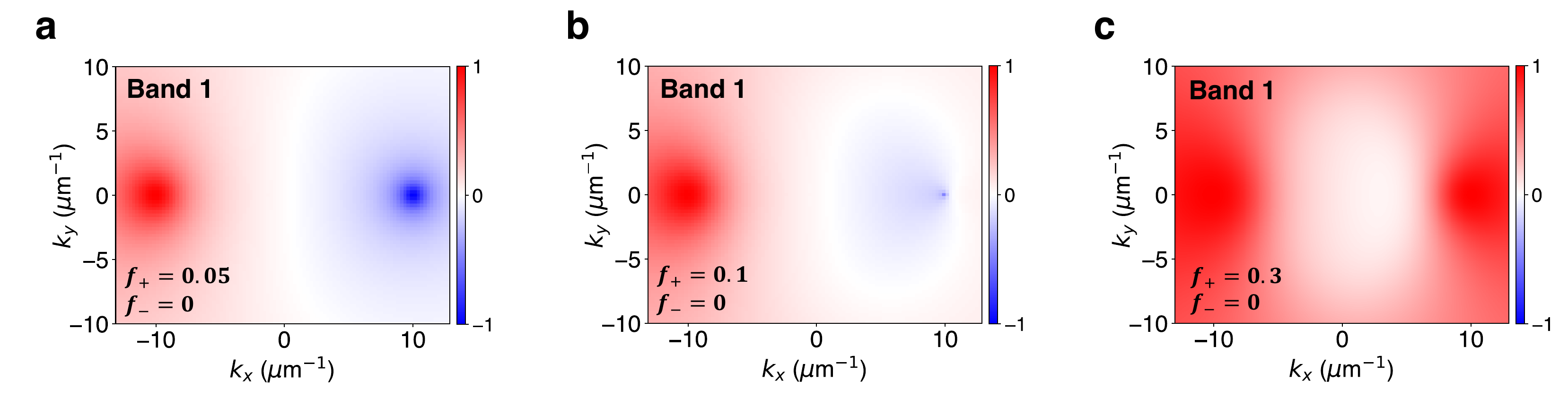}
		\caption{The Stokes parameter, $S_3(\mathbf{k})$, of the lowest energy band (Band 1) under pumping with $\sigma_+$ polarized light which creates populations (a) $f_+=0.05$, $f_-=0$, (b) $f_+=0.1$, $f_-=0$, and (c) $f_+=0.3$, $f_-=0$.}\label{fig:gradual}
	\end{figure}
	The number of excitations in the system $N_{\mathrm{ex}}=\sum_{\mathbf{k},\alpha}a^{\dagger}_{\mathbf{k},\alpha}a_{\mathbf{k},\alpha}+\sum_{\mathbf{n},\alpha}\sigma^{\dagger}_{\mathbf{n},\alpha}\sigma_{\mathbf{n},\alpha}$ is a conserved quantity of this Hamiltonian. Therefore, when we have $f_+$ fraction of molecules in the $\ket{+_{\mathrm{mol}}}$ state and $f_-$ in the $\ket{-_{\mathrm{mol}}}$ state, we will only have to look at the $(f_++f_-)N^{\mathrm{th}}$ excitation manifold. Unfortunately, the dimensions of the Hilbert space of this manifold scale as $\binom{N}{(f_++f_-)N}$, and this quickly becomes computationally intractable as the system size, $N$, increases. Using mean-field theory, we reduce this many-body problem to a one-body problem. That is, we derive an effective Hamiltonian for a single excitation in the mean-field of the remaining $(f_++f_-)N$ excitations; in this way, we reduce the dimensions of the Hilbert space to that of the first excitation manifold. To do this, we follow a procedure similar to that used by Ribeiro \textit{et al.} \cite{f2018theory} and write the Heisenberg equations of motion (EOM) for the operators $\hat{\sigma}_{\mathbf{m},\pm}$ and $\hat{a}_{\mathbf{k},\pm}$,
	\begin{equation}
		\begin{aligned}
			i\hbar\frac{d\hat{\sigma}_{\mathbf{n},\pm}}{dt}=&\comm{\hat{\sigma}_{\mathbf{n},\pm}}{\hat{H}_{\mathrm{mol}}}+\comm{\hat{\sigma}_{\mathbf{n},\pm}}{\hat{H}_{\mathrm{cav}}}+\comm{\hat{\sigma}_{\mathbf{n},\pm}}{\hat{H}_{\mathrm{cav-mol}}}\\
			=&\hbar\omega_{\mathrm{e}}\hat{\sigma}_{\mathbf{n},\pm}+\frac{1}{\sqrt{N_xN_y}}\sum_{\mathbf{k}}e^{i\mathbf{k}\cdot \mathbf{r_n}}\Big[(1-\hat{\sigma}_{\mathbf{n},\mp}^{\dagger}\hat{\sigma}_{\mathbf{n},\mp}-2\hat{\sigma}_{\mathbf{n},\pm}^{\dagger}\hat{\sigma}_{\mathbf{n},\pm})\Big(\mathbf{J}_{\mathbf{k},+}\cdot\bm{\mu}_{\pm}\hat{a}_{\mathbf{k},+}\\
			&+\mathbf{J}_{\mathbf{k},-}\cdot\bm{\mu}_{\pm}\hat{a}_{\mathbf{k},-}\Big)-\hat{\sigma}_{\mathbf{n},\mp}^{\dagger}\hat{\sigma}_{\mathbf{n},\pm}\Big(\mathbf{J}_{\mathbf{k},+}\cdot\bm{\mu}_{\mp}\hat{a}_{\mathbf{k},+}+\mathbf{J}_{\mathbf{k},-}\cdot\bm{\mu}_{\mp}\hat{a}_{\mathbf{k},-}\Big)\Big],\\
			i\hbar\frac{d\hat{a}_{\mathbf{k},\pm}}{dt}=&\comm{\hat{a}_{\mathbf{k},\pm}}{\hat{H}_{\mathrm{mol}}}+\comm{\hat{a}_{\mathbf{k},\pm}}{\hat{H}_{\mathrm{cav}}}+\comm{\hat{a}_{\mathbf{k},\pm}}{\hat{H}_{\mathrm{cav-mol}}}\\
			=&\Big(E_0+\frac{\hbar^2|\mathbf{k}|^2}{2m^*}\pm \zeta|\mathbf{k}|\cos\phi\Big)\hat{a}_{\mathbf{k},\pm}+\Big(-\beta_0+\beta|\mathbf{k}|^2e^{\mp i2\phi}\Big)\hat{a}_{\mp,\mathbf{k}}\\
			&+\frac{1}{\sqrt{N_xN_y}}\sum_{\mathbf{m}}e^{i\mathbf{k}\cdot \mathbf{r_m}}\Big(\mathbf{J}_{\mathbf{k},\pm}^*\cdot\bm{\mu}_{+}^*\hat{\sigma}_{\mathbf{m},+}+\mathbf{J}_{\mathbf{k},\pm}^*\cdot\bm{\mu}_{-}^*\hat{\sigma}_{\mathbf{m},-}\Big).
		\end{aligned}
	\end{equation}
	
	We make a mean-field approximation to linearize these EOM. For instance, we use $mn\approx\bar{m}n$, that is, 
	\begin{equation}
		\begin{aligned}
			\hat{\sigma}_{\mathbf{n},+}^{\dagger}\hat{\sigma}_{\mathbf{n},+}\hat{a}_{\mathbf{k},+}=&\Big(\langle\hat{\sigma}_{\mathbf{n},+}^{\dagger}\hat{\sigma}_{\mathbf{n},+}\rangle+	\hat{\sigma}_{\mathbf{n},+}^{\dagger}\hat{\sigma}_{\mathbf{n},+}-\langle\hat{\sigma}_{\mathbf{n},+}^{\dagger}\hat{\sigma}_{\mathbf{n},+}\rangle\Big)\hat{a}_{\mathbf{k},+}\\
			=&\langle\hat{\sigma}_{\mathbf{n},+}^{\dagger}\hat{\sigma}_{\mathbf{n},+}\rangle \hat{a}_{\mathbf{k},+}+(\hat{\sigma}_{\mathbf{n},+}^{\dagger}\hat{\sigma}_{\mathbf{n},+}-\langle\hat{\sigma}_{\mathbf{n},+}^{\dagger}\hat{\sigma}_{\mathbf{n},+}\rangle)\langle\hat{a}_{\mathbf{k},+}\rangle\\
			\approx& \langle\hat{\sigma}_{\mathbf{n},+}^{\dagger}\hat{\sigma}_{\mathbf{n},+}\rangle\hat{a}_{\mathbf{k},+},
		\end{aligned}
	\end{equation}
	where $\langle\hat{O}\rangle=\Tr[\hat{\rho}_0 \hat{O}]$ with $\hat{\rho}_0\approx\prod_{\mathbf{m}}\hat{\rho}_{\mathbf{m}}\prod_{\mathbf{k}}\prod_{\alpha=+,-}\hat{\rho}_{\alpha,\mathbf{k}}$ \cite{fowler2022efficient}. Here, we assume that after dephasing of the molecular amplitudes, $\hat{\rho}_{\mathbf{m}}=f_{\mathrm{G}}\ket{\mathbf{m},G}\bra{\mathbf{m},G}+f_{+}\ket{\mathbf{m},+_{\mathrm{mol}}}\bra{\mathbf{m},+_{\mathrm{mol}}}+f_{-}\ket{\mathbf{m},-_{\mathrm{mol}}}\bra{\mathbf{m},-_{\mathrm{mol}}}$, $\hat{\rho}_{\alpha,\mathbf{k}}=\ket{\mathbf{k},\alpha_{\mathrm{cav}},0}\bra{\mathbf{k},\alpha_{\mathrm{cav}},0}$, and, therefore, $\langle\hat{a}_{\mathbf{k},+}\rangle=0$. The EOM then become
	\begin{equation}
		\begin{aligned}
			i\hbar\frac{d\hat{\sigma}_{\mathbf{n},\pm}}{dt}
			\approx&\hbar\omega_{\mathrm{e}}\hat{\sigma}_{\mathbf{n},\pm}+\frac{1}{\sqrt{N_xN_y}}(1-f_{\mp}-2f_{\pm})\sum_{\mathbf{k}}e^{i\mathbf{k}\cdot \mathbf{r_n}}\Big(\mathbf{J}_{\mathbf{k},+}\cdot\bm{\mu}_{\pm}\hat{a}_{\mathbf{k},+}\\
			&+\mathbf{J}_{\mathbf{k},-}\cdot\bm{\mu}_{\pm}\hat{a}_{\mathbf{k},-}\Big),\\
			i\hbar\frac{d\hat{a}_{\mathbf{k},\pm}}{dt}
			=&\Big(E_0+\frac{\hbar^2|\mathbf{k}|^2}{2m^*}\pm \zeta|\mathbf{k}|\cos\phi\Big)\hat{a}_{\mathbf{k},\pm}+\Big(-\beta_0+\beta|\mathbf{k}|^2e^{\mp i2\phi}\Big)\hat{a}_{\mp,\mathbf{k}}\\
			&+\frac{1}{\sqrt{N_xN_y}}\sum_{\mathbf{m}}e^{i\mathbf{k}\cdot \mathbf{r_m}}\Big(\mathbf{J}_{\mathbf{k},\pm}^*\cdot\bm{\mu}_{+}^*\hat{\sigma}_{\mathbf{m},+}+\mathbf{J}_{\mathbf{k},\pm}^*\cdot\bm{\mu}_{-}^*\hat{\sigma}_{\mathbf{m},-}\Big).
		\end{aligned}
	\end{equation}    
	
	We define rescaled operators $\hat{\sigma}_{\mathbf{n},\pm}'=\hat{\sigma}_{\mathbf{n},\pm}/\sqrt{1-f_{\mp}-2f_{\pm}}$ and rewrite the EOM,
	\begin{equation}
		\begin{aligned}
			i\hbar\frac{d\hat{\sigma}_{\mathbf{n},\pm}'}{dt}
			\approx&\hbar\omega_{\mathrm{e}}\hat{\sigma}_{\mathbf{n},\pm}'+\frac{1}{\sqrt{N_xN_y}}\sqrt{1-f_{\mp}-2f_{\pm}}\sum_{\mathbf{k}}e^{i\mathbf{k}\cdot \mathbf{r_n}}\Big(\mathbf{J}_{\mathbf{k},+}\cdot\bm{\mu}_{\pm}\hat{a}_{\mathbf{k},+}\\
			&+\mathbf{J}_{\mathbf{k},-}\cdot\bm{\mu}_{\pm}\hat{a}_{\mathbf{k},-}\Big),\\
			i\hbar\frac{d\hat{a}_{\mathbf{k},\pm}}{dt}
			=&\Big(E_0+\frac{\hbar^2|\mathbf{k}|^2}{2m^*}\pm \zeta|\mathbf{k}|\cos\phi\Big)\hat{a}_{\mathbf{k},\pm}+\Big(-\beta_0+\beta|\mathbf{k}|^2e^{\mp i2\phi}\Big)\hat{a}_{\mp,\mathbf{k}}\\
			&+\frac{}{\sqrt{N_xN_y}}\sum_{\mathbf{m}}e^{i\mathbf{k}\cdot \mathbf{r_m}}\Big(\sqrt{1-f_--2f_+}\mathbf{J}_{\mathbf{k},\pm}^*\cdot\bm{\mu}_{+}^*\hat{\sigma}_{\mathbf{m},+}'+\sqrt{1-f_+-2f_-}\mathbf{J}_{\mathbf{k},\pm}^*\cdot\bm{\mu}_{-}^*\hat{\sigma}_{\mathbf{m},-}'\Big).
		\end{aligned}
	\end{equation}   
	From these EOM, along with the fact that $\hat{\sigma}_{\mathbf{n},\pm}'$ act effectively as bosonic operators in mean-field, $\comm{\hat{\sigma}_{\mathbf{n},+}'}{\hat{\sigma}^{\prime\dagger}_{\mathbf{n},+}}
	=\frac{1-\hat{\sigma}_{\mathbf{n},-}^{\dagger}\hat{\sigma}_{\mathbf{n},-}-2\hat{\sigma}_{\mathbf{n},+}^{\dagger}\hat{\sigma}_{\mathbf{n},+}}{1-f_--2f_+}\approx \hat{I}$ and
	$\comm{\hat{\sigma}_{\mathbf{n},+}'}{\hat{\sigma}^{\prime\dagger}_{\mathbf{n},-}}=\frac{-\hat{\sigma}_{\mathbf{n},-}^{\dagger}\hat{\sigma}_{\mathbf{n},+}}{1-f_--2f_+}\approx \hat{0}$, where $\hat{I}$ and $\hat{0}$ are the identity and zero operators, we can construct an effective Hamiltonian $\hat{H}^{\mathrm{eff}}=\hat{H}_{\mathrm{mol}}^{\mathrm{eff}}+\hat{H}_{\mathrm{cav}}^{\mathrm{eff}}+\hat{H}_{\mathrm{cav-mol}}^{\mathrm{eff}}$ in $\hat{\sigma}_{\mathbf{n},\pm}'$ and $\hat{a}_{\mathbf{k},\pm}$,
	\begin{equation}\label{eq:PosPump}
		\begin{aligned}
			\hat{H}_{\mathrm{mol}}^{\mathrm{eff}}=&\sum_{\mathbf{n}}\Big(\hbar\omega_{\mathrm{e}}\hat{\sigma}_{\mathbf{n},+}^{\prime\dagger}\hat{\sigma}_{\mathbf{n},+}'+\hbar\omega_{\mathrm{e}}\hat{\sigma}_{\mathbf{n},-}^{\prime\dagger}\hat{\sigma}_{\mathbf{n},-}'\Big),\\
			\hat{H}_{\mathrm{cav}}^{\mathrm{eff}}=&\sum_{\mathbf{k}}\Big(E_0+\frac{\hbar^2|\mathbf{k}|^2}{2m^*}+\zeta|\mathbf{k}|\cos\phi\Big)\hat{a}_{\mathbf{k},+}^{\dagger}\hat{a}_{\mathbf{k},+}\\
			&+\Big(E_0+\frac{\hbar^2|\mathbf{k}|^2}{2m^*}-\zeta|\mathbf{k}|\cos\phi\Big)\hat{a}_{\mathbf{k},-}^{\dagger}\hat{a}_{\mathbf{k},-}+\Big(-\beta_0+\beta|\mathbf{k}|^2e^{-i2\phi}\Big)\hat{a}_{\mathbf{k},+}^{\dagger}\hat{a}_{\mathbf{k},-}\\
			&+\Big(-\beta_0+\beta|\mathbf{k}|^2e^{i2\phi}\Big)\hat{a}_{\mathbf{k},-}^{\dagger}\hat{a}_{\mathbf{k},+},\\
			\hat{H}_{\mathrm{cav-mol}}^{\mathrm{eff}}=&\frac{1}{\sqrt{N_xN_y}}\sum_{\mathbf{m}}\sum_{\mathbf{k}}e^{i\mathbf{k}\cdot \mathbf{r_m}}\Bigg[\sqrt{1-f_{-}-2f_{+}}\Big(\mathbf{J}_{\mathbf{k},+}\cdot \bm{\mu}_{+}\hat{\sigma}_{\mathbf{m},+}^{\prime\dagger}\hat{a}_{\mathbf{k},+}\\
			&+\mathbf{J}_{\mathbf{k},-}\cdot \bm{\mu}_{+}\hat{\sigma}_{\mathbf{m},+}^{\prime\dagger}\hat{a}_{\mathbf{k},-}\Big)+\sqrt{1-f_{+}-2f_{-}}\Big(\mathbf{J}_{\mathbf{k},+}\cdot \bm{\mu}_{-}\hat{\sigma}_{\mathbf{m},-}^{\prime\dagger}\hat{a}_{\mathbf{k},+}\\
			&+\mathbf{J}_{\mathbf{k},-}\cdot \bm{\mu}_{-}\hat{\sigma}_{\mathbf{m},-}^{\prime\dagger}\hat{a}_{\mathbf{k},-}\Big)\Bigg]+\mathrm{H.c.},
		\end{aligned}
	\end{equation}
	which is the mean-field Hamiltonian when the system has $f_+,f_-$ excitations. Writing this effective Hamiltonian in k-space,
	\begin{equation}
		\begin{aligned}
			\hat{H}_{\mathrm{mol}}^{\mathrm{eff}}=&\sum_{\mathbf{k}}\Big[\hbar\omega_{\mathrm{e}}\hat{\sigma}_{\mathbf{k},+}^{\prime\dagger}\hat{\sigma}_{\mathbf{k},+}'+\hbar\omega_{\mathrm{e}}\hat{\sigma}_{\mathbf{k},-}^{\prime\dagger}\hat{\sigma}_{\mathbf{k},-}'\Big],\\
			\hat{H}_{\mathrm{cav}}^{\mathrm{eff}}=&\sum_{\mathbf{k}}\Big(E_0+\frac{\hbar^2|\mathbf{k}|^2}{2m^*}+\zeta|\mathbf{k}|\cos\phi\Big)\hat{a}_{\mathbf{k},+}^{\dagger}\hat{a}_{\mathbf{k},+}+\Big(E_0+\frac{\hbar^2|\mathbf{k}|^2}{2m^*}-\zeta|\mathbf{k}|\cos\phi\Big)\hat{a}_{\mathbf{k},-}^{\dagger}\hat{a}_{\mathbf{k},-}\\
			&+\Big(-\beta_0+\beta|\mathbf{k}|^2e^{-i2\phi}\Big)\hat{a}_{\mathbf{k},+}^{\dagger}\hat{a}_{\mathbf{k},-}+\Big(-\beta_0+\beta|\mathbf{k}|^2e^{i2\phi}\Big)\hat{a}_{\mathbf{k},-}^{\dagger}\hat{a}_{\mathbf{k},+},\\
			\hat{H}_{\mathrm{cav-mol}}^{\mathrm{eff}}=&\sum_{\mathbf{k}}\Bigg[\sqrt{1-f_{-}-2f_{+}}\Big(\mathbf{J}_{\mathbf{k},+}\cdot \bm{\mu}_{+}\hat{\sigma}_{\mathbf{k},+}^{\prime\dagger}\hat{a}_{\mathbf{k},+}\\
			&+\mathbf{J}_{\mathbf{k},-}\cdot \bm{\mu}_{+}\hat{\sigma}_{\mathbf{k},+}^{\prime\dagger}\hat{a}_{\mathbf{k},-}\Big)+\sqrt{1-f_{+}-2f_{-}}\Big(\mathbf{J}_{\mathbf{k},+}\cdot \bm{\mu}_{-}\hat{\sigma}_{\mathbf{k},-}^{\prime\dagger}\hat{a}_{\mathbf{k},+}\\
			&+\mathbf{J}_{\mathbf{k},-}\cdot \bm{\mu}_{-}\hat{\sigma}_{\mathbf{k},-}^{\prime\dagger}\hat{a}_{\mathbf{k},-}\Big)\Bigg]+\mathrm{H.c.}
		\end{aligned}
	\end{equation}
	We define states $\ket{\mathbf{k},\pm_{\mathrm{mol}}}'$ and $\ket{\mathbf{k},\pm_{\mathrm{cav}}}'$ corresponding to operators $\hat{\sigma}_{\mathbf{k},\pm}^{\prime\dagger}$ and $\hat{a}_{\mathbf{k},\pm}^{\dagger}$, respectively. Writing the Hamiltonian $\hat{H}^{\mathrm{eff}}(\mathbf{k})=\bra{\mathbf{k}}\hat{H}^{\mathrm{eff}}\ket{\mathbf{k}}$ in the above basis we obtain,
	\begin{equation}\label{eq:pumpHamil}
		\hat{H}^{\mathrm{eff}}(\mathbf{k})=\hat{H}^{\mathrm{eff}}_{\mathrm{mol}}(\mathbf{k})+\hat{H}^{\mathrm{eff}}_{\mathrm{cav}}(\mathbf{k})+\hat{H}^{\mathrm{eff}}_{\mathrm{cav-mol}}(\mathbf{k}),
	\end{equation}
	where,
	\begin{equation}\label{eq:pumpHamilexpand}
		\begin{aligned}
			\hat{H}^{\mathrm{eff}}_{\mathrm{mol}}(\mathbf{k})=&\hbar\omega_{\mathrm{e}}\ket{+_{\mathrm{mol}}}'\bra{+_{\mathrm{mol}}}'+\hbar\omega_{\mathrm{e}}\ket{-_{\mathrm{mol}}}'\bra{-_{\mathrm{mol}}}',\\
			\hat{H}^{\mathrm{eff}}_{\mathrm{cav}}(\mathbf{k})=&\Big(E_0+\frac{\hbar^2|\mathbf{k}|^2}{2m^*}+\zeta|\mathbf{k}|\cos\phi\Big)\ket{+_{\mathrm{cav}}}'\bra{+_{\mathrm{cav}}}'+\Big(E_0+\frac{\hbar^2|\mathbf{k}|^2}{2m^*}-\zeta|\mathbf{k}|\cos\phi\Big)\ket{-_{\mathrm{cav}}}'\bra{-_{\mathrm{cav}}}'\\
			&+\Big(-\beta_0+\beta|\mathbf{k}|^2e^{-i2\phi}\Big)\ket{+_{\mathrm{cav}}}'\bra{-_{\mathrm{cav}}}'+\Big(-\beta_0+\beta|\mathbf{k}|^2e^{i2\phi}\Big)\ket{-_{\mathrm{cav}}}'\bra{+_{\mathrm{cav}}}',\\
			\hat{H}^{\mathrm{eff}}_{\mathrm{cav-mol}}(\mathbf{k})=&\mathbf{J}_{\mathbf{k},+}\cdot \Big(\sqrt{1-f_--2f_+}\bm{\mu}_{+}\ket{+_{\mathrm{mol}}}'+\sqrt{1-f_+-2f_-}\bm{\mu}_{-}\ket{-_{\mathrm{mol}}}'\Big)\bra{+_{\mathrm{cav}}}'\\
			&+\mathbf{J}_{\mathbf{k},-}\cdot \Big(\sqrt{1-f_--2f_+}\bm{\mu}_{+}\ket{+_{\mathrm{mol}}}'+\sqrt{1-f_+-2f_-}\bm{\mu}_{-}\ket{-_{\mathrm{mol}}}'\Big)\bra{-_{\mathrm{cav}}}'+\mathrm{H.c.}
		\end{aligned}
	\end{equation}
	
	Upon pumping with circularly polarized light, the lowest band gradually changes from containing equal number of modes of both circular polarizations to overwhelmingly containing modes of a single polarization as $|f_+-f_-|$ increases (Fig. \ref{fig:gradual}).
	\section{Parameters}
	\subsection*{Perylene filled cavity}
	We take parameters for the perylene filled cavity $\beta_0 = 0.1 \mathrm{eV}$, $\beta = 9\times10^{-4} \mathrm{eV}\mu \mathrm{m}^2$, $\zeta = 2.5\times10^{-3}\mathrm{eV}\mu \mathrm{m}$, $m^* =125 \hbar^2\mathrm{eV}^{-1}\mu \mathrm{m}^{-2}$, and $L_z = 0.745 \mu \mathrm{m}$,
	where these are similar to those used to model the experiments of Ren \textit{et al.} \cite{ren2021nontrivial} (Fig. 3, 4, and 5 in main manuscript). On the other hand, we modify $E_0$ and $n_z$ such that they make the photon modes in our model near resonant with the transition that is strongly coupled to the cavity. For instance, we take $E_0=3.80\mathrm{eV}$ and $n_z=11$ for porphyrin (Fig. 3 and 4); $E_0=2.50 \mathrm{eV}$ and $n_z=9$ for Ce:YAG (Fig. 5b-c); and $E_0=1.80 \mathrm{eV}$ and $n_z=5$ for MoS$_2$ (Fig. 5e-f). We assume that perylene has a similar effect on these different photon modes, as it does on modes with $E_0\sim2.27$eV at $\mathbf{k}=0$ in experiments \cite{ren2021nontrivial}. This may not necessarily be true, however, as we consider a perylene filled cavity only to achieve frequency separation of photon modes with different polarization, and this can instead be easily achieved with an electrically tunable liquid crystal cavity \cite{rechcinska2019engineering}, replacing a perylene filled cavity with a liquid-crystal cavity will not modify the underlying physics of the phenomenon we are interested in, \textit{i.e.}, the idea of using saturation to break TRS will remain intact.
	
	\subsection*{Porphyrin, Ce:YAG, and monolayer MoS$_2$}
	We take areal density $\rho_A=3.55\times10^{5}\mu\mathrm{m}^{-2}$ ($\sim2000$ molecules in 75nm $\cross$ 75nm) \cite{hulsken2007real}, relative permittivity $\varepsilon=1.5$ \cite{li1993porphyrin}, frequency $\hbar\omega_{\mathrm{e}}=3.8056$eV and transition dipole $\mu_0=1.1184\mathrm{au}\times 2.5417 \mathrm{D}/\mathrm{au}=2.84\mathrm{D}$ \cite{sun2022polariton} for the porphyrin film. Also, we consider $100$ such porphyrin films stacked one over the other along the $z$ direction within the cavity to achieve strong light-matter coupling, $N_z=100$. Therefore, the effective areal density of molecules $\rho_A'=N_z\rho_A$ will be used instead of $\rho_A$ while computing $\mathbf{J}_{\mathbf{k},\alpha}$. These are the parameters used to generate Fig. 3 and 4.
	
	Similarly, using density $\rho_{\mathrm{YAG}}=5.11$g cm$^{-3}$, molar mass $M_{\mathrm{YAG}}=738$ g mol$^{-1}$, number of Y$^{3+}$ per unit cell $n_{\mathrm{Y}^{3+}}=3$, and concentration of Ce$^{3+}$ (relative to Y$^{3+}$) $1\%=10^{-2}$ \cite{bachmann2009temperature}, we obtain the effective areal density of Ce$^{3+}$ ions in a $L_z'=0.1\mu$m thick layer of Ce:YAG to be $\rho_A'=10^{-2}L_z'n_{\mathrm{Y}^{3+}}\rho_{\mathrm{YAG}} N_A/M_{\mathrm{YAG}}=1.25\times10^7\mu$m$^{-2}$. This will be used while computing $\mathbf{J}_{\mathbf{k},\alpha}$ in place of $\rho_A$. We use relative permittivity $\varepsilon=12$ \cite{ctibor2021dielectric} and frequency $\hbar\omega_{\mathrm{e}}=2.53$eV (489nm \cite{kolesov2013mapping}) for the transition in a Ce:YAG crystal. Using the oscillator strength of this transition $0.286$ \cite{kolesov2013mapping}, we calculate the transition dipole $\mu_0=5.46$D. These are the parameters used to generate Fig. 5c.
	
	For monolayer MoS$_2$, we consider A-excitons at $\hbar\omega_{\mathrm{e}}=1.855$eV \cite{chen2017valley}. From Chen \textit{et al}. \cite{chen2017valley}, we take the Rabi splitting at resonance, and use $\mu_0\sqrt{\rho_A}\sqrt{\hbar \omega_{\mathrm{e}}/2L_z\varepsilon\epsilon_0}\approx39\mathrm{meV}/2=19.5\mathrm{meV}$ in our calculations (Fig. 5f).

\renewcommand{\bibname}{References}
\bibliographystyle{myformat.bst}
\bibliography{supplementary}